\documentclass[11pt]{article}
\usepackage[margin=1in]{geometry}

\usepackage{dsfont}
\usepackage{amsfonts}
\usepackage{amsmath}
\usepackage{amssymb}
\usepackage{amsthm}
\usepackage{makeidx}
\usepackage{graphicx}
\usepackage{tabularx}
\usepackage{hyperref}

\usepackage{mathrsfs}
\usepackage{url}
\usepackage{stmaryrd}
\usepackage[labelformat=simple]{subcaption}

\usepackage{natbib}
\def\NatBibNumeric{%
}
 \bibpunct[, ]{(}{)}{,}{a}{}{,}%
 \def\bibfont{\small}%
 \def\newblock{\ }%

\usepackage{color}
\definecolor{orange}{RGB}{250, 54, 0}
\definecolor{purple}{rgb}{0.75, 0.0, 1.0}

\usepackage[normalem]{ulem}

\newcommand{\RR}{\mathbb{R}}

\newcommand{\CN}{\mathcal{N}}
\newcommand{\BB}{\boldsymbol\Delta}
\newcommand{\BBs}{\overline{\boldsymbol\Delta}}
\newcommand{\DB}{\mathcal{D}_{\Delta} p(\Pi) }
\newcommand{\BP}{\boldsymbol\Pi}
\newcommand{\BA}{\boldsymbol\Pi^n}
\newcommand{\DE}{\mathcal{D}_{\vec{E}(\Pi)}p(\Pi)}

\newcommand{\lb}{\left\lbrace}
\newcommand{\rb}{\right\rbrace}
\newcommand{\pbar}{\bar{p}}
\newcommand{\one}{\mathds{1}}

\newcommand{\beq}{\begin{equation}}
\newcommand{\eeq}{\end{equation}}

\newtheorem{thm}{Theorem}[section]
\newtheorem{prop}[thm]{Proposition}
\newtheorem{rem}[thm]{Remark}
\newtheorem{coro}[thm]{Corollary}
\newcommand{\bprof}{\begin{proof}}
\newcommand{\eprof}{\end{proof}}
\newcommand{\bprop}{\begin{prop}}
\newcommand{\eprop}{\end{prop}}

\newtheorem{dfn}[thm]{Definition}
\newcommand{\bdfn}{\begin{dfn}}
\newcommand{\edfn}{\end{dfn}}

\newtheorem{exmpl}[thm]{Example}
\newcommand{\bexmpl}{\begin{exmpl}}
\newcommand{\eexmpl}{\end{exmpl}}

\title{Sensitivity of the Eisenberg--Noe clearing vector to individual interbank liabilities.\footnote{The views expressed in this work are those of the authors and do not necessarily reflect the views of Norges Bank. 
This material is based upon work supported by the National Science Foundation under Grant No.\ 1321794. }}
\date{}
\author{
 Zachary Feinstein\thanks{Washington University in St.\ Louis, Department of Electrical \& Systems Engineering, 1 Brookings Drive, Green Hall, Room 2160B, St.\ Louis, MO 63130, USA, {\tt zfeinstein@ese.wustl.edu}.}, 
	Weijie Pang\thanks{Worcester Polytechnic Institute, Department of Mathematical Sciences, 100 Institute Road, Worcester, MA 01609-2280 USA, {\tt wpang@wpi.edu} and {\tt ssturm@wpi.edu}.},
	Birgit Rudloff\thanks{Vienna University of Economics and Business, Welthandelsplatz 1, Building D4, 1020 Vienna, Austria, {\tt brudloff@wu.ac.at}.}, \\
	Eric Schaanning\thanks{Norges Bank Research, Bankplassen 2, 0107 Oslo, Norway, {\tt eric.schaanning@norges-bank.no}.}~~\thanks{RiskLab, ETH Z\"{u}rich, Department of Mathematics, R\"{a}mistrasse 101, 8092 Z\"{u}rich. Part of this work was undertaken while Eric Schaanning was at Imperial College London.},
	Stephan Sturm\footnotemark[3], 
	Mackenzie Wildman\thanks{University of California Santa Barbara, Department of Statistics \& Applied Probability,  South Hall, Santa Barbara, CA 93106-3080. Part of this work was undertaken while Mackenzie Wildman was at Lehigh University, {\tt mackenzie.wildman@gmail.com}.}
}

\begin{document}

\maketitle
\begin{abstract}
\noindent
We quantify the sensitivity of the Eisenberg--Noe clearing vector to estimation errors in the bilateral liabilities of a financial system. 
The interbank liabilities matrix is a crucial input to the computation of the clearing vector. 
However, in practice central bankers and regulators must often estimate this matrix because complete information on bilateral liabilities is rarely available.
As a result, the clearing vector may suffer from estimation errors in the liabilities matrix. 
We quantify the clearing vector's sensitivity to such estimation errors and show that its directional derivatives are, like the clearing vector itself, solutions of fixed point equations. 
We describe estimation errors utilizing a basis for the space of matrices  representing permissible perturbations and derive analytical solutions to the maximal deviations of the Eisenberg--Noe clearing vector.  
This allows us to compute upper bounds for the worst case perturbations of the clearing vector. 
Moreover, we quantify the probability of observing clearing vector deviations of a certain magnitude, for uniformly or normally distributed errors in the relative liability matrix.

Applying our methodology to a dataset of European banks, we find that perturbations to the relative liabilities can result in economically sizeable differences that could lead to an underestimation of the risk of contagion.
Our results are a first step towards allowing regulators to quantify errors in their simulations.
\\

\end{abstract}

\clearpage

\section{Introduction}


Some important streams of the literature on contagion in networks has focused on interbank contagion, building on the network model of \cite{EisenbergNoe:2001}. 
Central banks and regulators have applied the model to study default cascades in their jurisdictions' banking systems. (\cite{Anand:Stresstesting_Canadian_banking_system2014,HalajKok:ModelingEmergenceInterbank:2015,Boss/Elsinger/Summer/Thurner:2004,ElsingerLeharSummer:NetworkModelsAndSystemicRisk:2013,Upper:2011,Gai2011complexity}). 
{\cite{Hueser_review:2016} provides a comprehensive and detailed review of the interbank contagion literature. 
\cite{hurd2016contagion} presents a unified mathematical framework for modeling these contagion channels. }
Recently, the Bank of England has extended this model to analyse solvency contagion in the UK financial system (\cite{Brinley:Decline_of_Solvency_contagion2017}).
{
Multiple, extensions of this model have been developed to include effects such as 
\begin{itemize}
\item \textit{Bankruptcy costs}: \cite{Elsinger:09,rogers_failure_2013,ElliottGolubJackson:14,GlassermanYoung:ContagionFinancialNetworks2014,AwiszusWeber:JointImpact2015},
\item \textit{Cross-ownershi}p:  \cite{Elsinger:09,ElliottGolubJackson:14,AwiszusWeber:JointImpact2015}
\item \textit{Fire sales}:  \cite{CifuentesFerruciShin:LiquidityRiskandContagion2005,NierYangYorulmazerAlentorn,GaiKapadia,
ChenLiuYao:ModelingFinancialSystemicRisk_NetworkEffectMarketLiquidityEffect2014,AminiFilipovicMinca:CCP,
AminiFilipovicMinca:FireSale,AwiszusWeber:JointImpact2015,Feinstein:MultipleIlliquid,FeinsteinElMasri:Leverage,
Feinstein:Multilayered,Lillo:SystemicRisk_Entropy}, and
\item \textit{Multiple maturities}:  \cite{CapponiChen:Dynamic,veraartKusnetsov,Feinstein:Dynamic}. 
\end{itemize} 
} 
{Moreover, a number of papers analyze the implications of network topology on systemic risk in greater detail.
\cite{amini2016resilience} derive rigorous asymptotic results for the magnitude of the default cascade in terms of network characteristics and find that institutions that have large connectivity and a high number of ``contagious links" contribute most to contagion. 
\cite{detering2016managing} show that if the degree distribution of the network does not have a second moment, local shocks can propagate through the entire network. This is relevant as realistic financial networks typically display a core-periphery structure with inhomogeneous degree distribution (\cite{Cont/Moussa/Santos:2012}). 
\cite{chong2018contagion} characterise the joint default distribution of a financial system for all possible network structures and show how Bayesian network theory can be applied to detect contagious channels. 
}

Regulators have identified the inclusion of such contagion mechanisms in stress tests as a key priority  (\cite{BCBS:making_supervisory_stress_tests_macroprudential}, \cite{LSE:transatlantic_assessment}).
Furthermore, recent research illustrates that accounting for feedback effects and contagion can change the pass/fail result in stress tests for individual institutions (\cite{empiricalpaper}). 

A key ingredient required to estimate contagion in these models is the so-called liabilities matrix $L$, where $L_{ij}$ is the nominal liability of bank $i$ to bank $j$. 
Often, the exact bilateral exposures are not known and thus need to be estimated (\cite{HalajKok:ECB,Anand:Filling_in_the_blanks:2015,ElsingerLeharSummer:NetworkModelsAndSystemicRisk:2013,HalajKok:ModelingEmergenceInterbank:2015}). 
Despite considerable efforts after the crisis to improve data collection, data gaps have not been closed yet.
Beyond logistical issues like the standardization of reporting formats and the creation of unique and universal institution identifiers, further hurdles remain, such as legal restrictions that limit regulators' access only to data pertinent to their respective jurisdictions. 
Therefore, the estimation of specific bilateral exposures remains an important issue (\cite{Langfield:mapping_UK:2014,Anand:Filling_in_the_blanks:2015,Anand:slides2015,FSB_IMF:progress_report2015}). 
The early literature often used entropy maximizing techniques to ``fill in the blanks" in the liabilities matrix given the total assets and liabilities of banks (viz. the row and column sums of $L$). 
However, a growing empirical literature has shown that real-world interbank networks look quite different from the homogeneous networks that are obtained with such techniques (\cite{Bech:topology_federal_funds_marekt:2010,Mistrulli:2011,Cont/Moussa/Santos:2012,Soramaki:interbank_payment_flows2007}). 
A recent Bayesian method to estimate the bilateral liabilities, given the total liabilities and potential other prior information,  is proposed in \cite{gandy_bayesian_2015} and applied to reconstruct CDS markets in \cite{Gandy_veraart:adjustable_network_reconstruction_2017}. 
In particular, \cite{Mistrulli:2011,gandy_bayesian_2015} show how wide estimates of systemic risk may fluctuate when estimating contagion on real-world and heterogeneous networks versus uniform networks.  
This highlights the pivotal role that the matrix of bilateral exposures plays in quantifying the extent of contagion when computing default cascades. 
Beyond the above-mentioned legal hurdles that restrict regulator's access to data outside their jurisdiction, another important example of uncertainty in the interbank exposures arises due to time gaps between data collection and the run of the stress test: For some regulatory stress tests (e.g.\ Dodd-Frank stress tests) data is collected annually, which can both give rise to window-dressing behaviour by banks, as well as exposures naturally changing over time. In this case the existence or non-existence of an exposure between two banks will be known, and the uncertainty mainly surrounds its magnitude. 
\cite{CapponiChenYao} studies the effects of the network topology on systemic risk through the use of majorization-based tools.
To the best of our knowledge, \cite{liu_sensitivity_2010} is so far the only paper that performs a sensitivity analysis of the Eisenberg--Noe model. 
Their analysis 
focuses on
the sensitivity of the clearing vector with respect to the initial net worth of each bank. \\ \\
%
%
The main contribution of this paper is to perform a detailed sensitivity analysis of the clearing vector with respect to the interbank liabilities in the standard Eisenberg--Noe framework. 
To this end, we define directional derivatives of order $k$ of the clearing vector with respect to ``perturbation matrices," which quantify the estimation errors in the relative liability matrix. 
This allows us to derive an exact Taylor series for the clearing vector. 
Moreover, we introduce a set of ``basis matrices," which specify a notion of  fundamental directions for the directional derivative. We demonstrate that the directional derivative of the clearing vector can be written as a linear combination of these basis matrices. 
We proceed to use this result to study two optimisation problems that quantify the maximal deviation of the clearing vector from its ``true" value, and obtain explicit solutions for both problems. 
These analytical results additionally provide an upper bound to the (first-order) worst case perturbation error. We extend these results by computing the probability of observing deviations of a given magnitude when the estimation errors are either uniformly or normally distributed.

Finally, we illustrate our results both in a small four-bank network and using a dataset of European banks. 
Our results suggest that, though the set of defaulting banks may remain stable across different bilateral interbank networks (calibrated to the same data set), the deviation of the clearing vector from perturbations in the relative liabilities can be large.  
While our stylized setting ignores other extensions of the Eisenberg--Noe framework (such as bankruptcy costs or fire sales for instance), it provides a first step towards quantifying the sensitivity of the clearing vector to the liabilities matrix, which has not been addressed in the literature before.\\ \\
In this paper, we occasionally consider external liabilities along with the interbank liabilities.  We aggregate all external liabilities into a single external ``societal firm.''  This additional ``bank'' is a stand-in for the entirety of the economy that is not included in the financial network.  This is discussed in more details in, e.g., \cite{GlassermanYoung:ContagionFinancialNetworks2014}.  In particular, as utilized in \cite{feinstein_measures_2015}, the impact on the wealth of the societal firm can be used as an aggregate measure for the health of the financial network as a whole.  We will make use of the societal firm in a similar way in order to study the effects of estimation errors in the interbank liabilities on external stakeholders.

{We have limited the literature review mainly to papers that are close to the Eisenberg--Noe methodology. Needless to say, since the financial crisis a vast number of papers have been written on measuring systemic risk, using different approaches such as Agent-Based Modeling (\cite{Bookstaber:ABMFinancialVulnerability}), Econometry (\cite{SRISK}), Mean-Field Games (\cite{carmona2013mean}) or Economic analysis (\cite{brunnermeier2014measuring, Hellwig2009}).
Axiomatic measures of systemic risk and set-theory approaches have been developed in (\cite{chen2013axiomatic, kromer2016systemic, biagini2015unified, feinstein_measures_2015}). 
\cite{bisias}, \cite{FouqueLangsam:Handbook_systemic_risk} and \cite{Duffie:banksfail} provide broad overviews of this vast literature. } \\ \\
The organization of the paper is as follows.  In Section~\ref{sec:NetworkModel} we present the Eisenberg--Noe framework and 
provide initial continuity results of that model.  We then study  directional derivatives and the Taylor series of the Eisenberg--Noe clearing payments with respect to the relative liabilities matrix.  These results allow us to consider the sensitivity of the clearing payments.  In Section~\ref{sec:perturbation} we use the directional derivatives in order to determine the perturbations to the relative liabilities matrix that present the ``worst'' errors in terms of misspecification of the clearing payments and impact to society.  These results are extended to also consider the probability of the various estimation errors.  In Section~\ref{sec:EBA} we implement our sensitivity analysis on data calibrated to a network of European banks. Section~\ref{sec:conc} concludes {with a summary and a discussion of the limitations of our approach}. Technical proofs are mostly relegated to the appendix which also provides details on the orthogonal basis of perturbation matrices.

\section{Sensitivity analysis of Eisenberg--Noe clearing vector}\label{sec:NetworkModel}
We consider a financial system consisting of $n$  banks, $\CN = \{ 1, \ldots, n \}$. 
For $i,j  \in \CN$, $L_{ij} \geq 0$ is the \textbf{nominal liability} of bank $i$ to bank $j$.\footnote{External liabilities can be considered as well through the introduction of an ``external'' bank $0$.  This is discussed in more detail in Section~\ref{sec:society}.}
Equivalently, $L_{ij}$ is the exposure of bank $j$ to bank $i$. 
 $L \in \RR^{n\times n}$ is called the \textbf{liabilities matrix} of the financial network, and we assume that no bank has an exposure to itself, i.e., $L_{ii} = 0$ for all $i \in \CN$.
The \textbf{total liability} of bank $i$ is given by $\pbar_{i} = \sum_{j = 1}^{n} L_{ij}$. The \textbf{relative liability} of bank $i$ to bank $j$ is denoted $\pi_{ij}\in[0,1]$, where $\pi_{ij} = \frac{L_{ij}}{\pbar_{i}}$ when $\pbar_{i}>0$. We allow $\pi_{ij}\in[0,1]$ to be arbitrary when $\pbar_i=0$ and only require $\sum_{j=1}^n \pi_{ij}=1$.\footnote{Note that the arbitrary choice of $\pi_{ij}$ in the case $\pbar_i=0$ has no impact on the outcome of the Eisenberg--Noe model since the transpose of the relative liability matrix $\Pi$ is multiplied by the incoming payment vector $p(\Pi)$, whose $j^{\text{th}}$ entry is 0 when $\pbar_j=0$ (cf.\ \eqref{def:pBar}).}
We denote the \textbf{relative liability matrix} $\Pi \in \RR^{n\times n}$. 
Any relative liability matrix $\Pi$ must belong to the set of admissible matrices $\BA$, defined as the set of all right stochastic matrices with entries in $[0,1]$ and all diagonal entries 0:
\begin{equation}\label{def:BA}
\BA := \biggl\{ \Pi\in [0,1]^{n\times n} \; \Big| \; \forall i: \pi_{ii} = 0 ,\;  \sum_{j=1}^n \pi_{ij} = 1 \biggr\}.
\end{equation}
Finally, denote the \textbf{external assets} of bank $i$ from outside the banking system by $x_{i} \geq 0$.
A bank balance sheet then takes the simplified form of Table \ref{table:balance_sheet}, and a financial system is given by the triplet $(\Pi, x, \bar{p}) \in \BA \times \RR_{+}^{n} \times \RR_{+}^{n}$. 

\begin{table}
\centering
\begin{tabular}{|c|c||c|c|}
\hline
\textbf{Assets}	& Representation & \textbf{Liabilities}	&  Representation \\
\hline
\hline
Interbank  &  $a_{i}^{IB} = \sum_{j = 1}^{n} L_{ji}$  & 
Interbank  & $l_{i}^{IB} = \sum_{j=1}^n L_{ij}  $\\
External  & $x_{i}$ &  External  &   $L_{i0}  $ \\ 
 &  &  Capital &   $c_{i}  $  \\
 \hline   
\end{tabular}
\caption{ Stylized bank balance sheet }\label{table:balance_sheet}
\end{table}

\begin{figure}
\centering
\begin{subfigure}{.45\textwidth}
	\centering
	\includegraphics[width=\linewidth]{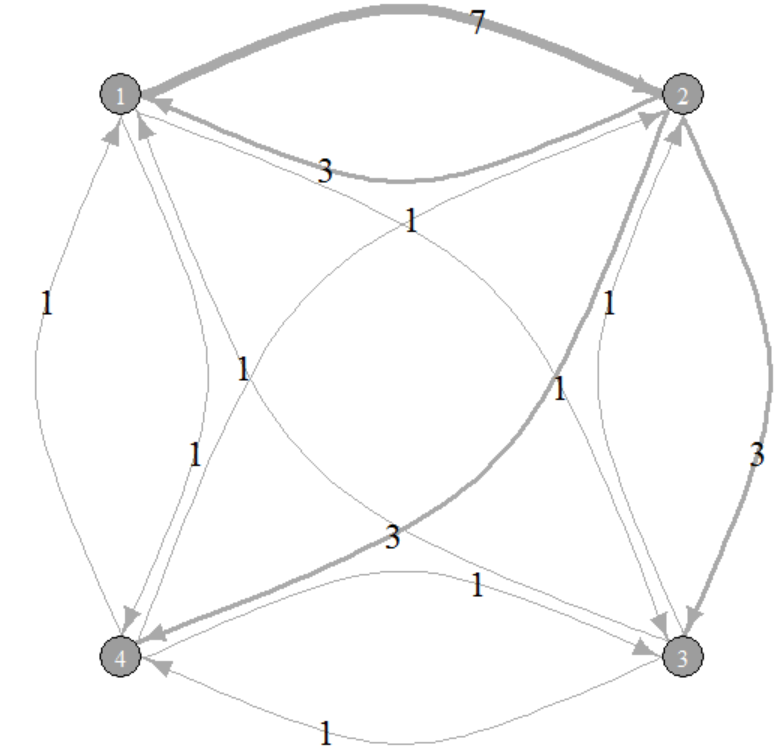}
	\caption{Nominal interbank liabilities}\label{fig:nominal_liabilties}
\end{subfigure}
~
\begin{subfigure}{.45\textwidth}
	\centering
	\includegraphics[width=\linewidth]{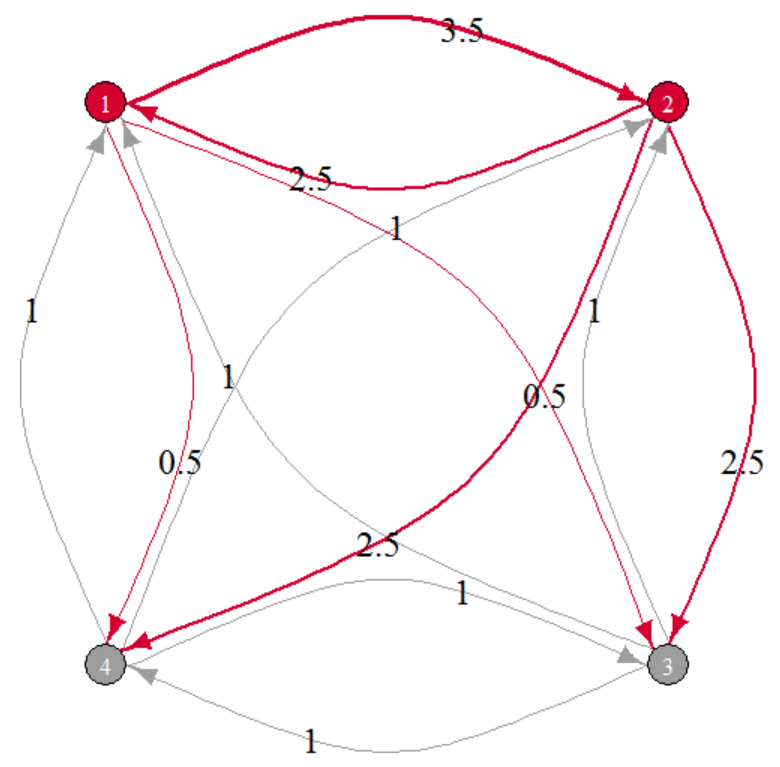}
	\caption{Clearing interbank payments}\label{fig:clearing_payments}
\end{subfigure}
\caption{Initial network defined in Example \ref{ex:clearing_payments}}
\end{figure}

A bank is solvent when the sum of its net external assets and performing interbank assets exceeds its total liabilities.  
In this case, the bank honours all of its obligations. 
However, if the value of its obligations is greater than the bank's net assets plus performing interbank assets, then the bank will default and repay its obligations pro-rata.
\footnote{This corresponds to the assumption that all interbank and external claims can be aggregated to a single figure per bank and that all creditors of a defaulting bank are paid pari passu.} 
These rules yield a \textbf{clearing vector} as the solution of the fixed point problem
\begin{equation}\label{def:pBar}
p(\Pi) = \pbar \wedge \bigl( x + \Pi^{\top} p(\Pi) \bigr) . 
\end{equation}
Let $p: \BA \to \RR_{+}^{n} ; 
\Pi \mapsto p(\Pi)$ be the fixed point function with parameters $(x,\pbar)$. 
As proved in \cite[Theorem 2]{EisenbergNoe:2001}, the clearing vector is unique if a system of banks is regular. Regularity is defined as follows: 
A \textbf{surplus set} $S \subseteq \CN$ is a set of banks in which no bank in the set has any obligations to a bank outside of the set and the sum over all banks' external net asset values in the set is positive, i.e., $\forall \, (i,j) \in  S \times S^{c}:  \pi_{ij} = 0$ and $\sum_{i \in S} x_{i} > 0$. Next, consider the financial system as a directed graph in which there is a directed link from bank $i$ to bank $j$ if $L_{ij} > 0$. Denote the \textbf{risk orbit} of bank $i$ as 
$o(i) = \{j \in \CN  \mid \text{ there exists a directed path from } i \text{ to } j\}$. 
This means that the risk orbit of bank $i$ is the set of all banks which may be affected by the default of bank $i$.
A system is \textbf{regular} if every risk orbit is a surplus set.
Uniqueness of the clearing vector has important consequences in terms of the continuity of the function $p$, which in turn is important for our sensitivity analysis. 
For this reason we will proceed under the assumption that our financial system is regular.  
\begin{prop}\label{prop:continuity}
Consider a regular financial system $(\Pi, x, \bar{p})$ in which $x$ and $\pbar$ are fixed. The function $p$, defined via \eqref{def:pBar},
	is continuous with respect to $\Pi \in \BA$.
\end{prop}
We finish these preliminary notes by considering a simple example of the Eisenberg--Noe clearing payments under a system of $n = 4$ banks.  We will return to this example throughout as a simple illustrative case study.
\bexmpl\label{ex:clearing_payments}
Consider the following example of a network consisting of four banks in which the bank's nominal interbank liabilities are given by 
\begin{equation*}
L = 
\begin{pmatrix}
0 & 7 & 1 & 1 \\
3 & 0 & 3 & 3  \\
1 & 1 & 0 & 1 \\
1 & 1 & 1 & 0 \\
\end{pmatrix},
\end{equation*} 
as shown in Figure \ref{fig:nominal_liabilties}. Assume the banks' external assets are given by the vector $x=(0,2,2,2)^\top$. With 0 net worth and positive liabilities, Bank 1 defaults initially. 
The Eisenberg--Noe clearing vector \eqref{def:pBar} can be easily computed to be $p = (4.5, 7.5, 3, 3)^{\top}$, showing that Bank 2 also defaults through contagion. The realized interbank payments are shown in Figure \ref{fig:clearing_payments}. Banks who are in default are colored red and payments that are repaid less than whole are also colored red. The edge widths are proportional to the payment size.

\eexmpl

\subsection{Quantifying estimation errors from the (relative) liabilities matrices}
We assume that some estimation error is attached to the entries of the relative liability matrix, leading to a deviation of the clearing vector from the ``true" clearing vector $p(\Pi)$. 
Denote the true relative liabilities matrix by $\Pi$ and let $\Pi + h \Delta$ denote the liabilities matrix that includes some estimation error, for a perturbation matrix $\Delta$ and size $h \in \RR$. 
First we consider the class of perturbation matrices, $\BB^n(\Pi)$, under which we assume that the existence or non-existence of a link between two banks is known to the regulator and hence, the error is limited to a misspecification of the size of that link. 
In practice, this type of uncertainty arises when data is collected at a low frequency, which can lead to exposure evolving naturally, as well as banks trying to improve their balance sheet composition ahead of regulatory reporting dates.\footnote{
Evidence for such behaviour at the end of a quarter can, for instance, be seen in the balance sheet reduction of European Banks and the corresponding spikes this creates in the utilization of the Federal Reserve's Reverse Repo facility, see: 
\url{http://libertystreeteconomics.newyorkfed.org/2017/08/regulatory-incentives-and-quarter-end-dynamics-in-the-repo-market.html}.
}

Remark \ref{rem:29}, Corollary \ref{cor:worst_case} and Corollary \ref{cor:worst_case_society} will utilize the results in this section to provide bounds for the perturbation error in general without predetermining existence or non-existence of links.  
\bdfn\label{def:B} For a fixed $\bar{p} \in \RR^n_+ $, define the set of relative liability perturbation matrices by
\begin{equation*}
\BB^{n} (\Pi) := \biggl\lbrace 
\Delta \in \RR^{n \times n}  \; \biggm| \; \forall  i:  \delta_{ii} = 0,~  \sum_{j=1}^n \delta_{ij} = 0,~  \sum_{j=1}^n \delta_{ji} \bar{p}_j  = 0  ,\text{ and } ( \pi_{ij}=0 ) \Rightarrow ( \delta_{ij}=0)\, \forall \, j 
\biggr\rbrace . 
\end{equation*}
\edfn
\noindent The summation conditions ensure that the total liabilities and total assets, respectively, of each bank are left unchanged by the perturbation. 
Of course it is not possible to have $\Pi + h \Delta \in \BA$ for any $h\in\RR$.
Throughout this work we consider perturbation magnitudes in a bounded interval, $h \in ( -h^{*},h^{*} )$, where
\begin{equation*}
h^{*} 
:= \min \left\lbrace  \min_{\delta_{ij} < 0, \, \, \pbar_i>0} \frac{- \pi_{ij}}{\delta_{ij}} \, ,  \, \min_{\delta_{ij} > 0, \, \, \pbar_i>0}  \frac{1 - \pi_{ij}}{\delta_{ij}} \right\rbrace > 0,
\end{equation*}
for any $\Delta \in \BB^n(\Pi)$ to assure $\Pi + h \Delta \in \BA$.
We exclude from this calculation of $h^*$ any bank $i$ where $\pbar_i=0$ since this has no impact on the results.
It is natural to consider directional derivatives on a unit-ball, whence we focus on the bounded set of perturbations
\begin{equation*}
\BB_{F}^{n} (\Pi ) := \BB^{n} (\Pi)  \cap \lb \Delta \in \RR^{n \times n} \; \big| \; \| \Delta \|_{F} \leq 1 \rb  ,
\end{equation*}
where $\|\cdot\|_F$ is the Frobenius norm, i.e., $\|\Delta\|_F = \sqrt{\sum_{i = 1}^n \sum_{j = 1}^n |\delta_{ij}|^2}$.

\begin{rem}
A more general case can be considered in which one allows for errors that create links where there were none or remove connections where there was one. 
This set is defined as follows: For a fixed $\bar{p} \in \RR^n_+ $,
\begin{equation*}
\BBs^{n}(\Pi) := \biggl\lbrace 
 \Delta \in \RR^{n \times n} \; \biggm| \; \forall  i:  \delta_{ii} = 0,~  \sum_{j=1}^n \delta_{ij} = 0  ,~  \sum_{j=1}^n \delta_{ji} \bar{p}_j  = 0 , \text{ and } ( \pi_{ij}=0 ) \Rightarrow ( \delta_{ij} \geq 0 )  \, \forall \, j
\biggr\rbrace . 
\end{equation*}
We will consider in particular the bounded set of perturbations
\begin{equation*}
\BBs^{n}_{F}(\Pi) := \BBs^n(\Pi) \cap \lb \Delta \in \RR^{n \times n} \; \big| \; \| \Delta \|_{F} \leq 1 \rb .
\end{equation*}
{Such perturbations thus allow a ``rewiring" of the network. In general, allowing edges to be added or deleted increases the potential error in the clearing vector. 
However, the infinitesimal nature of the sensitivity analysis necessarily restricts the rewiring to the creation of new links; any strictly positive liability cannot be deleted through an infinitesimal perturbation.
We discuss this issue in more detail in Corollary \ref{cor:worst_case}, where we apply our methodology to the complete network, as well as in Figure \ref{fig:EBA_worst_estimation_error}, which shows a distribution of payouts to society under a rewiring of the interbank network. }
\end{rem}

\subsection{Directional derivatives of the Eisenberg--Noe clearing vector} 
Next, we analyse the error when using the clearing vector of a perturbed liability matrix, $p(\Pi + h \Delta)$, instead of the clearing vector of the original liability matrix, $p(\Pi)$, for small perturbations $h \Delta$, with $\Delta \in \BB^n(\Pi)$. 
\bdfn\label{def:directional derivative}
Let $\Delta \in \BB^n(\Pi)$. In the case that the following limit exists, we define the directional derivative of the clearing vector $p(\Pi)$ in the direction of a perturbation matrix $\Delta$ as
\begin{equation*}
\DB :=   \lim_{h \to 0} \frac{p(\Pi+h \Delta)  - p(\Pi ) }{ h }. 
\end{equation*}
\edfn
\noindent The first order Taylor expansion of $p$ about $\Pi$ gives
\begin{equation*}
	p(\Pi+h \Delta) - p(\Pi) = h \DB+ O\bigl(h^2\bigr). 
\end{equation*}
The following theorem provides an explicit formula for the directional derivative of the clearing vector for a fixed financial network.
\begin{thm}\label{thm:derivative}
Let $(\Pi, x, \bar{p})$ be a regular financial system. The directional derivative of the clearing vector $p(\Pi)$ in the direction of a perturbation matrix $\Delta \in \BB^n(\Pi)$ exists almost everywhere and is given by
\begin{equation}\label{eqn:directional derivative}
\DB= \Bigl(  I - \text{diag}(d) \Pi^{\top}  \Bigr)^{-1}\text{diag}(d) \Delta^{\top}  p(\Pi) , 
\end{equation}
where $\text{diag}(d)$ is the diagonal matrix defined as
$
 \text{diag} (d_{1}, \ldots, d_{n}),  
$
where 
\begin{equation*}
d_{i} := \one_{\{x_i + \sum_{j=1}^n \pi_{ji}p_j(\Pi) < \pbar_i\}}.
\end{equation*}
Here, \eqref{eqn:directional derivative} holds outside of the measure-zero set $\lbrace x \in \RR_{+}^{n} \mid \exists i \in \CN \text{ s.t. } x_{i} + \sum_{j = 1}^n \pi_{ji}p_j(\Pi) $
$= \pbar_{i} \rbrace$ in which some bank is exactly at the brink of default.
\end{thm}
The term $(I - \text{diag} (d) \Pi^{\top} )^{-1}$ also appears in  \cite{ChenLiuYao:ModelingFinancialSystemicRisk_NetworkEffectMarketLiquidityEffect2014}, which the authors call the ``network multiplier." 
This multiplier appears in the dual formulation of the linear program characterising the Eisenberg--Noe clearing vector, where the authors introduce it to study the sensitivities of the clearing vector with respect to the capital (of defaulting banks) and the total liabilities (of non-defaulting banks).
The computation of the directional derivative above can be viewed as a generalisation of this result to arbitrary perturbations.
The interpretation remains the same in our case: The ``network multiplier" describes how an estimation error propagates through the network.

\subsection{A Taylor series for the Eisenberg--Noe clearing payments}
In the same manner, we can define higher order directional derivatives. 
\bdfn
For $k\geq1$, we define the $k^{th}$ order directional derivative of the clearing vector with respect to a perturbation matrix $\Delta$ as
\begin{equation}\label{eqn:higher order derivative}
\mathcal{D}_{\Delta}^{(k)} p(\Pi) := \lim_{h\rightarrow 0} \frac{\mathcal{D}_{\Delta}^{(k-1)} p(\Pi+h\Delta)  - \mathcal{D}_{\Delta}^{(k-1)} p(\Pi) }{h},
\end{equation}
when the limit exists,
and
\begin{equation*}
\mathcal{D}_{\Delta}^{(0)} p(\Pi) = p(\Pi).
\end{equation*}
\edfn
\noindent Remarkably, as Theorem \ref{thm:higher_order_derivatives} shows, all higher order derivatives also have an explicit formula, which allows us to obtain an exact Taylor series for the clearing vector. We impose an additional assumption on allowable perturbations $h\Delta$ so that the matrix $\text{diag}(d)$ (as defined in Theorem \ref{thm:derivative}) as a function of $\Pi+h\Delta$ is fixed with respect to $h$, i.e., we require $h$ sufficiently small so that the same subset of banks is in default when the liability matrix is $\Pi+h\Delta$ as when the liability matrix is $\Pi$. Let
\begin{align}
\nonumber \overline{h}^{**}
	&:= \sup\left\{h \leq h^* \; \left| \; \begin{array}{l}
		x_i+\sum_{j=1}^n\pi_{ji}p_j(\Pi)<\bar{p}_i \\ 
		\qquad \Leftrightarrow \; x_i+\sum_{j=1}^n (\pi_{ji} + h \delta_{ji} ) p_j(\Pi+h\Delta)<\bar{p}_i 
		\;\; \forall i\in\CN
		\end{array}\right.\right\},\\
\nonumber \underline{h}^{**} 
	&:= \inf\left\{h \geq -h^* \; \left| \; \begin{array}{l}
		x_i+\sum_{j=1}^n\pi_{ji}p_j(\Pi)<\bar{p}_i \\ 
		\qquad \Leftrightarrow \; x_i+\sum_{j=1}^n (\pi_{ji} + h \delta_{ji} ) p_j(\Pi+h\Delta)<\bar{p}_i 
		\;\; \forall i\in\CN
		\end{array}\right.\right\},\\
\label{eqn:hstarstar} h^{**} &:= \min\{-\underline{h}^{**},\overline{h}^{**}\}.
\end{align}
We necessarily have $h^{**} > 0$ because we exclude the measure-zero set 
$\lbrace x \in \RR_{+}^{n} \mid \exists i \in \CN \text{ s.t. } x_{i} + \sum_{j = 1}^n \pi_{ji} p_j(\Pi) = \pbar_{i} \rbrace$
in which a bank is exactly at the brink of default.
\begin{thm}\label{thm:higher_order_derivatives}
Let $(\Pi, x, \bar{p})$ be a regular financial system. Then for $\Delta \in \BB^{n}(\Pi)$, and 
for all $k \geq 1$:
\begin{align}\label{eq:nderivative}
\mathcal{D}_{\Delta}^{(k)} p(\Pi) & = k \bigl(I - \text{diag}(d) \Pi^{\top} \bigr)^{-1} \text{diag}(d) \Delta^{\top} \mathcal{D}_{\Delta}^{(k-1)} p(\Pi) \\
\nonumber & = k! \Bigl( \bigl( I - \text{diag}(d) \Pi^{\top} \bigr)^{-1} \text{diag}(d) \Delta^{\top} \Bigr)^{k} p(\Pi), 
\end{align}
where $\mathcal{D}_{\Delta}^{(0)} p(\Pi) = p(\Pi)$.
Moreover, for $h \in (-h^{**},h^{**})$, the Taylor series  
\begin{equation}\label{eqn:taylor_series}
p( \Pi + h \Delta) = \sum_{k = 0}^\infty \frac{h^k}{k!} \mathcal{D}_{\Delta}^{(k)} p(\Pi) 
\end{equation}
converges and has the following representation
\begin{equation}\label{eq:taylor}
p(\Pi + h \Delta ) = \Bigl(I - h \bigl( I - \text{diag}(d) \Pi^{\top} \bigr)^{-1} \text{diag}(d) \Delta^{\top} \Bigr)^{-1} p(\Pi)
\end{equation}
outside of the measure-zero set
$\lbrace x \in \RR_{+}^{n} \mid \exists i \in \CN \text{ s.t. } x_{i} + \sum_{j = 1}^n \pi_{ji} p_j(\Pi) = \pbar_{i} \rbrace$.
\end{thm}

Comparing the directional derivative \eqref{eqn:directional derivative} to the full Taylor series \eqref{eqn:taylor_series} allows us to make the interpretation of the ``network multiplier" more precise: 
The network multiplier captures the \textit{first order} effect of the error propagation in the final ``\textit{round}" of the fictitious default algorithm. The \textit{$k^{th}$ order} effect of the error propagation is captured by the network multiplier raised to the $k^{th}$ power.  Finally, the Taylor series of the fixed-point is the infinite series of these $k^{th}$ order network multipliers; as this is of a similar form it can be interpreted as the multiplier of the network multiplier. 

\begin{rem}\label{rem:29}
We can extend the Taylor series expansion results to the more general space of perturbation matrices $\BBs^n(\Pi)$ rather than $\BB^n(\Pi)$.  Over such a domain the Taylor series \eqref{eq:taylor} is only guaranteed to converge for 
\[h \in \biggl[0,\min\biggl\{\overline{h}^{**}, \frac{1}{\rho\bigl(( I - \text{diag}(d) \Pi^{\top} )^{-1} \text{diag}(d) \Delta^{\top}\bigr)}\biggr\}\biggr),\] 
as negative perturbations are not feasible.
\end{rem}

\begin{figure}
\centering
\includegraphics[width = 0.49\textwidth]{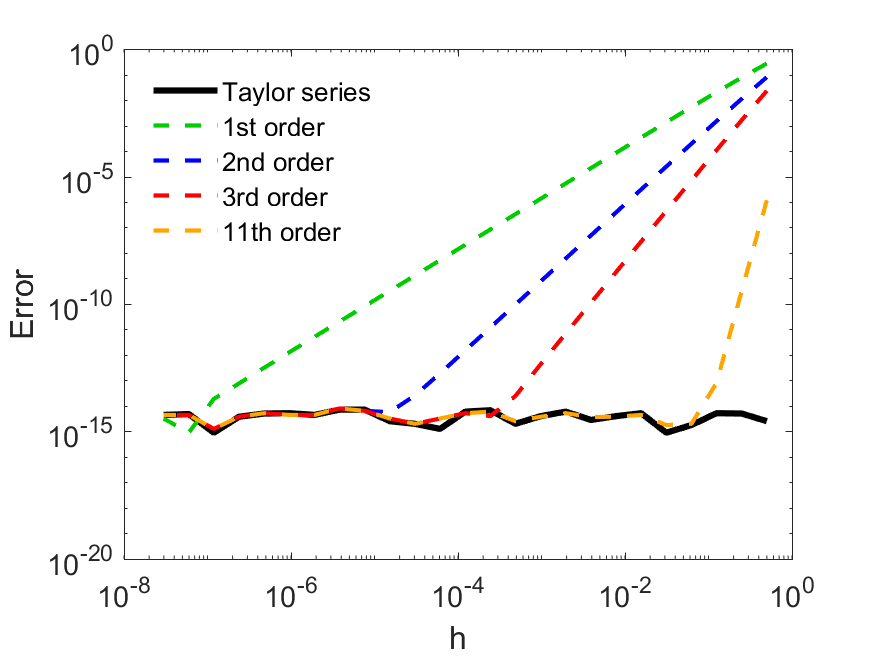}
\caption{Loglog plot of the approximation error, $|| p(\Pi) - p (\Pi + h \Delta) ||_{2}$ against the size of the perturbation $h$ for a random perturbation of the network introduced in Example 2.2.}
\end{figure}
\section{Perturbation errors}\label{sec:perturbation}
In this section we study in detail estimation errors in an Eisenberg--Noe framework, relying on the directional derivatives discussed in the previous section. Specifically we calculate both maximal errors as well as the error distribution assuming a specific distribution of the mis-estimation of the interbank liabilities, notably uniform and Gaussian. We do this first in the original Eisenberg--Noe model, considering the Euclidean norm of the clearing vector as objective. Then we turn to an enhanced model that includes an additional node representing society and study the effect of estimation errors on the payout to society.
\subsection{Deviations of the clearing vector}
We concentrate first on the $L^2$-deviation of the actual clearing vector from the estimated one.
	
\subsubsection{Largest shift of the clearing vector}\label{sec:worst_case}
We return to the first order directional derivative to quantify 
the largest shift of the clearing vector for estimation errors in the relative liability matrix given by perturbations in $\BB^n(\Pi)$. 
Let $\Delta \in \BB^{n} (\Pi)$  and assume that for a given $h \in \RR: \Pi + h \Delta \in \BA$. 
Then, the worst case estimation error under $\BB^n(\Pi)$ is given as
\begin{equation*}
\max_{\Delta \in \BB^{n} (\Pi)}  \|p(\Pi + h \Delta) - p(\Pi)\|_2^2. 
\end{equation*}
In order to remove the dependence on $h$ and the magnitude of $\Delta$, we consider instead the bounded set of directions $\BB_{F}^n(\Pi)$ and infintesimal perturbations,
\begin{equation*}
\max_{\Delta \in \BB_{F}^{n} (\Pi )} \lim_{h \to 0} \frac{\|p(\Pi + h \Delta) - p(\Pi)\|_2^2}{ h^{2}} = \max_{\Delta \in \BB_{F}^{n} (\Pi )}~  \|\DB\|_2^2.
\end{equation*}

In this section, we call $\|\DB\|_2^2$ the estimation error and $\max_{\Delta \in \BB^{n}_{F} (\Pi)} \|\DB\|_2^2$ the maximal deviation in the clearing vector under $\BB_{F}^n(\Pi)$.
Because $\Delta$ appears via a linear term in \eqref{eqn:directional derivative},  this allows us to use a basis of perturbation matrices in an elegant way to quantify the deviation of the Eisenberg--Noe clearing vector under the space of perturbations $\BB_F^n(\Pi)$. 

Throughout the following results we will take advantage of an orthonormal basis $\vec{E}(\Pi) = (E_1,\ldots,E_d)$ of the space $\BB^n(\Pi)$.  More details of this space are given in Appendix \ref{sec:batrices}.
\begin{prop}\label{prop:worst_case}
Let $(\Pi, x , \pbar)$ be a regular financial system. 
The worst case first order estimation error under $\BB_F^n(\Pi)$ is given by 
\begin{align}\label{eq:worst_case}
    \max_{\Delta \in \BB_{F}^{n}(\Pi)} \| \DB\|_2^2 = \bigl(\|\DE\|_2^o\bigr)^2
    \end{align}
    for any choice of basis $\vec{E}(\Pi)$ where 
$\|\cdot\|_2^o$ denotes the spectral norm of a matrix.    
 Furthermore, the largest shift of the clearing vector is achieved by 
\begin{equation*}
    \Delta^{*} (\Pi) : = \pm\sum_{k=1}^{d} z_k E_k,
\end{equation*}
where $z_{k}$ are the components of the (normalised) eigenvector corresponding to the maximum eigenvalue of $\bigl(\DE\bigr)^{\top} \DE$.
\end{prop}
\proof{}
Note first that any perturbation matrix $\Delta \in \BB^n(\Pi)$ can be written as a linear combination of basic perturbation matrices, i.e., $\Delta = \sum_{k=1}^{d} z_k E_k$.  Thus,
\begin{equation*}
    \| \DB\|_2^2 
    = \biggl\|\sum_{k=1}^{d} z_k \mathcal{D}_{E_k}p(\Pi)\biggr\|_2^2 
    = z^{\top} \bigl(\DE\bigr)^{\top} \DE z.
\end{equation*}
Immediately this implies, denoting the largest eigenvalue of a matrix $A$ by $\lambda_{\max}(A)$,
\begin{align*}
    \max_{\Delta \in \BB_{F}^{n}(\Pi)} \| \DB\|_2^2 & 
    = \max_{\|z \|_2\leq 1} z^{\top} \bigl(\DE\bigr)^{\top} \DE z \\
&    = \lambda_{\max} \Bigl(\bigl(\DE\bigr)^{\top} \DE\Bigr) \\
 &   = \bigl(\|\DE\|_2^o\bigr)^2.
\end{align*}
Finally, the independence of the solution from the choice of basis $\vec{E}(\Pi)$ is a direct result of Proposition \ref{prop:basis_eigenvalue}. 
\endproof
{Hence, if the true liability matrix were perturbed in the direction of $\Delta^{*} (\Pi)$, this would generate the largest first order estimation error in the clearing vector. By error, we mean the Euclidean distance between the ``true" clearing vector in the standard Eisenberg--Noe framework, and the clearing vector under the perturbed liabilities matrix. This is in general not equivalent to the direction that would change the default set most rapidly. 
Moreover, if regulatory expert judgement allowed to estimate reasonable absolute perturbations, our infinitesimal methodology could be used iteratively in a greedy approach until such an absolute estimation error was reached. }

We can use this result on the maximum deviations of the clearing vector under $\BB_F^n(\Pi)$ in order to provide bounds of the worst case perturbation error without predetermining the existence or non-existence of links.
\begin{coro}\label{cor:worst_case} 
Let $(\Pi, x , \pbar)$ be a regular financial system. 
The worst case first order estimation error under all perturbations is bounded by 
\begin{equation}\label{eq:worst_case_UB}
\bigl(\|\DE\|_2^o\bigr)^2 \leq \max_{\Delta \in \BBs_{F}^{n} (\Pi) } \| \DB \|_{2}^2 \leq \bigl(\| \mathcal{D}_{\vec{E}(\Pi_C)}p(\Pi) \|_{2}^{o}\bigr)^2
\end{equation}
for any choice of orthonormal bases $\vec{E}(\Pi)$ as above and $\vec{E}(\Pi_C)$ of any completely connected network $\Pi_C$. In the case that $\Pi$ itself is a completely connected network then this upper bound is attained.
\end{coro}
\proof{}
For all $\Pi$ and all completely connected networks $\Pi_C$, we have $\BB_F^n(\Pi) \subseteq \BBs_{F}^{n} (\Pi) \subseteq \BB_{F}^{n} (\Pi_C)$.
Hence, using~\eqref{eqn:directional derivative}, one obtains 
\begin{align*}
\max_{\Delta \in \BBs_{F}^{n} (\Pi)} \| \DB \|_{2}^2
& \leq \max_{\Delta \in \BB_{F}^{n} (\Pi_C) } \| \DB \|_{2}^2 \\
& = \max_{\|z\|_2 \leq 1} \biggl\|
(I - \text{diag}(d)\Pi^{\top})^{-1} \text{diag}(d)\biggl[ \sum_{k = 1}^d z_{k} E_{k} \biggr]^{\top} p(\Pi)  \biggr\|_{2}^2 \\
& = \bigl(\| \mathcal{D}_{\vec{E}(\Pi_C)}p(\Pi) \|_{2}^{o}\bigr)^2,\\
\max_{\Delta \in \BBs_{F}^{n} (\Pi)} \| \DB \|_{2}^2
& \geq \max_{\Delta \in \BB_{F}^{n} (\Pi) } \| \DB \|_{2}^2\\
& = \bigl(\|\DE\|_2^o\bigr)^2,
\end{align*}
where $\vec{E}(\Pi_C) := (E_{1}, \ldots, E_{d})$ is an orthonormal basis of the space $\BB^{n} (\Pi_C)$. 
As in Proposition \ref{prop:worst_case}, the independence of the solution from the choice of basis $\vec{E}(\Pi_C)$ is a direct result of Proposition \ref{prop:basis_eigenvalue}. 
\endproof

\begin{rem}
Our empirical analysis suggests that this bound is quite sharp (see Figure \ref{fig:EBA_society}).
\end{rem}

\begin{figure}
\centering
	\includegraphics[width=.45\linewidth]{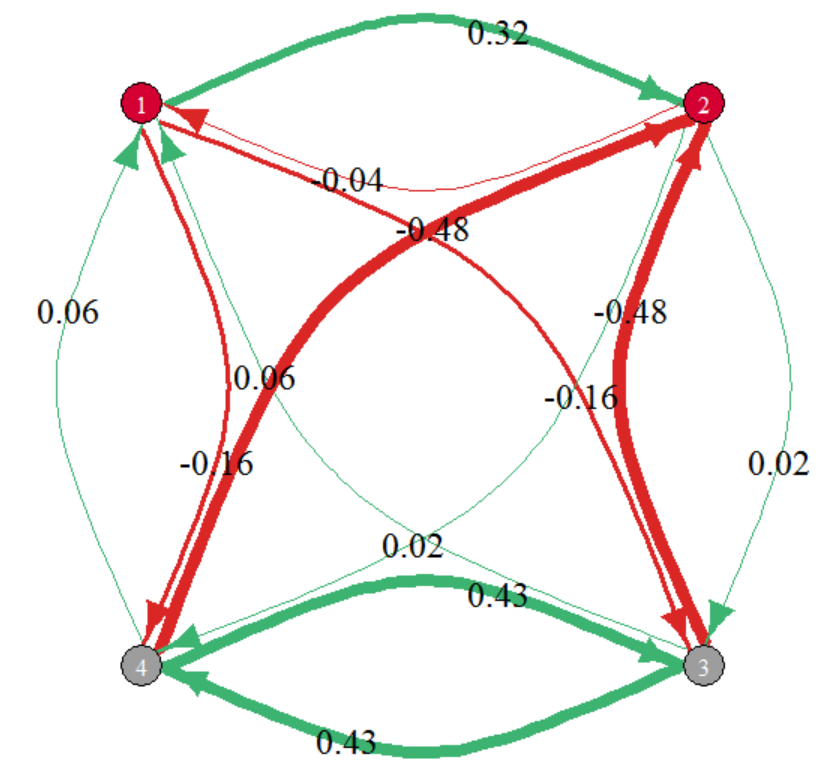}
	\caption{Worst case network perturbation under $\BB^n_F(\Pi)$ defined in Example \ref{ex:worstcase}}\label{fig:worst_case}
\end{figure}

\bexmpl\label{ex:worstcase}
We return to Example \ref{ex:clearing_payments} and consider the same toy network consisting of four banks in which each bank's nominal liabilities are shown in Figure \ref{fig:nominal_liabilties}. The largest shift of the clearing vector \eqref{eq:worst_case} under $\BB_F^4(\Pi)$, as described in Proposition \ref{prop:worst_case}, is given by the matrix
\begin{equation*}
    \Delta^*(\Pi) = 
\begin{pmatrix}
0 & 0.3230 & -0.1615 & -0.1615 \\
-0.0381 & 0 & 0.0190 & 0.0190 \\
0.0571& - 0.4845 & 0 & 0.4274 \\
0.0571 & -0.4845 & 0.4274 & 0
\end{pmatrix}.
\end{equation*}
As this network is complete, this is furthermore a solution to both optimization problems \eqref{eq:worst_case} and \eqref{eq:worst_case_UB} for the worst case perturbation.  Additionally, the upper bound in Corollary \ref{cor:worst_case} is attained.
This perturbation is depicted in Figure \ref{fig:worst_case}. As before, banks who are in default are colored red. The edges are labeled with the perturbation of the respective link between banks that achieves this greatest estimation error. The edge linking one node to another is red if the greatest estimation error under the set of perturbations $\BB_F^n(\Pi)$ occurs when we have overestimated the value of this link and green if we have underestimated it. Note that due to the symmetry of the optimal estimation error problem, $-\Delta^*(\Pi)$ is also optimal and thus the interpretation of red and green links in Figure \ref{fig:worst_case} can be reversed.  Indeed, when studying the deviation of the clearing vector, the solutions $\Delta^*(\Pi)$ and $-\Delta^*(\Pi)$ are equivalent. When analysing the shortfall of payments to society in Section \ref{sec:society}, this will be no longer the case. 
Edge widths are proportional to the absolute value of the entries in $\Delta^*(\Pi)$.
Though our Taylor expansion results (Theorem~\ref{thm:higher_order_derivatives}) are provided for $h \in (-h^{**},h^{**})$ only, the strict inequality is only necessary if $h^{**}$ denotes the perturbation size at which a new bank defaults, not when a connection is removed. 
So when $h = h^{**} \approx 0.688$, 
we obtain
\begin{equation*}
L^{*} = 
\begin{pmatrix}
0 & 9 & 0 & 0 \\
2.76 & 0 & 3.12 & 3.12 \\
1.12 & 0  & 0 &1.88\\
1.12 & 0  &1.88 & 0
\end{pmatrix}
,
\end{equation*} 
which has the clearing vector 
$$\hat{p} \approx ( 4.11, 6.11, 3, 3 )^{\top}.$$
One can immediately verify that $L^{*}$ has indeed the same total interbank assets and liabilities for each bank, but they are distributed in a different manner. 
Hence, in this example, there can be a deviation of up to 15\% in the relative norm of the clearing vector for a network that is still consistent with the total assets and total liabilities. 

\eexmpl

\begin{rem}
It may be desirable to normalize the first order estimation errors by, e.g., the clearing payments or total nominal liabilities, rather than considering the absolute error.  In a general form, let $A \in \RR^{n \times n}$ denote a normalization matrix (e.g., $A = \text{diag}(p(\Pi))^{-1}$ or $A = \text{diag}(\pbar)^{-1}$). Then we can extend the results of Proposition~\ref{prop:worst_case} and Corollary~\ref{cor:worst_case} by
\begin{align*}
\max_{\Delta \in \BB_{F}^{n}(\Pi)} \| A\DB\|_2^2 &= \bigl(\|A\DE\|_2^o\bigr)^2\\
\max_{\Delta \in \BBs_{F}^{n}(\Pi)} \| A\DB\|_2^2 &\leq \bigl(\|A \mathcal{D}_{\vec{E}(\Pi_C)}p(\Pi)\|_2^o\bigr)^2
\end{align*}
for any completely connected network $\Pi_C$.  Similarly the distribution results presented below can be generalized by considering $A\DE$ in place of $\DE$.
%
\end{rem}

\subsubsection{Clearing vector deviation for uniformly distributed estimation errors}
In this section, we will extend the above analysis to the case when estimation errors are uniformly distributed. 
This is done by considering the linear coefficients $z$ for the basis of perturbation matrices to be chosen uniformly on the $d$-dimensional Euclidean unit ball.  Then $\Delta = \sum_{k = 1}^{d}z_k E_k$ is a perturbation matrix.

\begin{prop}\label{prop:wce-uniform}
Let $(\Pi, x , \pbar)$ be a regular financial system. 
The distribution of the estimation error when the perturbations are uniformly distributed in the $L^2$-unit ball is given by 
\begin{align*}
    \mathds{P}&\bigl(\|\DB\|_2^2 \leq \alpha\bigr)	=	\frac{\text{vol}\bigl(\bigl\{ w \in \RR^{d}    
    \,\big|\, w^\top w \leq 1, w^\top \Lambda w \leq \alpha  \bigr\}\bigr)\Gamma\bigl(\frac{{d}}{2}+1\bigr)}{\pi^{{d}/2}}, \quad \alpha \geq 0,
\end{align*}
where $\Lambda$ is the diagonal matrix with elements given by the eigenvalues of $\bigl(\DE\bigr)^{\top} \DE$ for any choice of orthonormal basis $\vec{E}(\Pi)$, $\text{vol}$ denotes the volume operator, and $\Gamma$ is the gamma function.
\end{prop}
\proof{}
			Let $z$ be uniform on the $d$-dimensional unit ball. Then $\Delta = \sum_{k=1}^{d}z_k E_k$ is a perturbation matrix.
			One obtains
			\begin{align*}
							\mathds{P}\bigl(\|\DB\|_2^2 \leq \alpha \bigr) 
					=&	\mathds{P}\Bigl( \bigl( \DB \bigr)^\top\DB \leq \alpha \Bigr) \\
					=&	\mathds{P}\Bigl(z^\top \bigl(\DE\bigr)^{\top} 
															\DE z \leq \alpha \Bigr). 
			\end{align*}
			The matrix $\bigl(\DE\bigr)^{\top} \DE$ is diagonalizable because it is
			real and symmetric. Therefore we can write
			\begin{equation*}
				\bigl(\DE\bigr)^{\top} \DE = V^\top \Lambda V,
			\end{equation*}
			where $\Lambda$ is a diagonal matrix of the eigenvalues and $V$ is orthonormal.
			Combining the above equations, we have
			\begin{align*}
							\mathds{P}\Bigl(z^\top \bigl(\DE\bigr)^{\top} 
															\DE z \leq \alpha \Bigr) 
					=&	\mathds{P}\bigl(z^\top V^\top \Lambda V z \leq \alpha \bigr). 
			\end{align*}
			Then since $z$ is uniform on the unit ball and $VV^\top=I$, $w=Vz$ is also uniform on the unit ball and thus we have
			\begin{align*}
							\mathds{P}\bigl(z^\top V^\top \Lambda V z \leq \alpha \bigr) 
					=&	\mathds{P}\bigl(w^\top \Lambda w \leq \alpha \bigr) \\
					=&	\frac{\text{vol}\bigl(\bigl\{ w \;\big|\; w^\top w \leq 1, w^\top \Lambda w \leq \alpha  \bigr\}\bigr)}
									{\text{vol}\bigl(\bigl\{ w \;\big|\; w^\top w \leq 1  \bigr\}\bigr)}\\
					=&	\frac{\text{vol}\bigl(\bigl\{ w \;\big|\; w^\top w \leq 1, w^\top \Lambda w \leq \alpha \bigr\}\bigr)\Gamma\bigl(\frac{{d}}{2}+1\bigr)}
                    {\pi^{{d}/2}}.
			\end{align*}
 		As in Proposition \ref{prop:worst_case}, the independence of the distribution from the choice of basis $\vec{E}(\Pi_C)$ is a direct result of Proposition \ref{prop:basis_eigenvalue}.        
\endproof
     
\begin{rem}
In the case where $\alpha \leq \min_k \lambda_k$ or $\alpha \geq \max_k \lambda_k$ then $\mathds{P}\bigl(\|\DB\|_2^2 \leq \alpha\bigr)$ can explicitly be given by $\alpha^{{d}}\prod_{k = 1}^d \frac{1}{\sqrt{\lambda_k}}$ and $1$ respectively where $\{\lambda_k \; | \; k = 1,\ldots,d\}$ is the collection of eigenvalues of $\bigl(\DE\bigr)^{\top} \DE$. 
In the case that $\min_k \lambda_k < \alpha < \max_k \lambda_k$, the probability $\mathds{P}\bigl(\|\DB\|_2^2 \leq \alpha\bigr)$ can be given via the volume formula provided in Proposition~\ref{prop:wce-uniform} as ${d}$ nested integrals, 
	\[\frac{\Gamma\bigl(\frac{{d}}{2}+1\bigr)}{\pi^{{d}/2}} \int_{-1}^1 \int_{-\sqrt{1-x_{1}^2}}^{\sqrt{1-x_{1}^2}} \cdots \int_{-\sqrt{1-\sum_{k = 1}^{m-1} x_{k}^2}}^{\sqrt{1-\sum_{k = 1}^{m-1} x_{k}^2}} \int_{-\sqrt{\frac{\alpha - \sum_{k = 1}^m \lambda_{[k]}x_{k}^2}{\lambda_{[m+1]}}}}^{\sqrt{\frac{\alpha - \sum_{k = 1}^m \lambda_{[k]}x_{k}^2}{\lambda_{[m+1]}}}} \cdots \int_{-\sqrt{\frac{\alpha - \sum_{k = 1}^{d-1} \lambda_{[k]}x_{k}^2}{\lambda_{[d]}}}}^{\sqrt{\frac{\alpha - \sum_{k = 1}^{{d}-1} \lambda_{[k]}x_{k}^2}{\lambda_{[{d}]}}}} dx_{d}\cdots dx_{1},\]
	where $\lambda_{[m]} \leq \alpha \leq \lambda_{[m+1]}$  and $\lambda_{[m]}$ is a reordering of the eigenvalues such that $0 \leq \lambda_{[1]} \leq \lambda_{[2]} \leq \cdots \leq \lambda_{[{d}]}$.
\end{rem}
	
\begin{figure}
\centering
\includegraphics[width=0.48\textwidth]{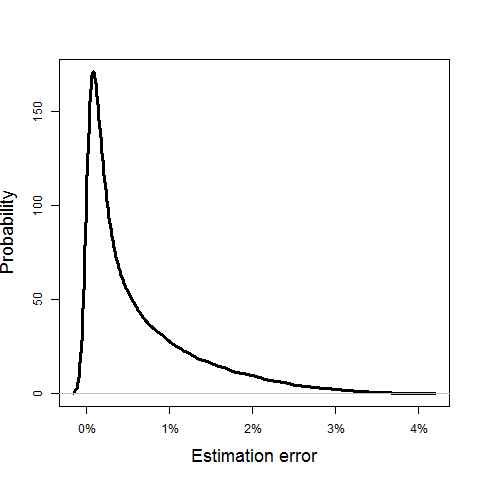}
\hspace{2ex}
\includegraphics[width=0.48\textwidth]{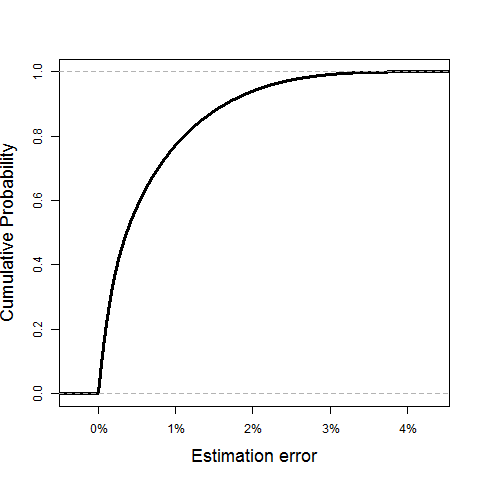}
\caption{The Probability density (left) and the CDF (right) of the relative estimation error, $\| \DB\|_2^2/\|p(\Pi)\|_2^2$, under uniform perturbations $\Delta$ as described in Example \ref{ex:uniform}}
\label{fig:prob_esterror_ball}
\end{figure}

\bexmpl	\label{ex:uniform}		
We return again to Example~\ref{ex:clearing_payments} to consider perturbations $\Delta$ sampled from the uniform distribution.  Figure \ref{fig:prob_esterror_ball} shows the density and CDF estimation for the relative estimation error, $\| \DB\|_2^2/\|p(\Pi)\|_2^2$, corresponding to our stylized four-bank network. The probabilities are estimated from 100,000 simulated uniform perturbations.
\eexmpl

\subsubsection{Clearing vector deviation for normally distributed estimation errors}\label{sec:worst_case_gaussian}
We extend our analysis from the previous subsection by considering normally distributed perturbations. To do so, we consider the linear coefficients $z$ for the basis of perturbation matrices to be chosen distributed according to  the standard ${d}$-dimensional multivariate standard Gaussian distribution.  Then $\sum_{k = 1}^{d}z_k E_k$ is a perturbation matrix $\Delta$. Though our prior results on the deviations of the clearing payments have been within the unit ball $\BB^n_F(\Pi)$, under a Gaussian distribution the magnitude of the perturbation matrices are no longer bounded by 1 and thus the estimation errors can surpass the worst case errors determined in Proposition \ref{prop:worst_case} and Corollary \ref{cor:worst_case}.
\begin{prop}\label{prop:wce-gaussian}
Let $(\Pi, x , \pbar)$ be a regular financial system. 
The distribution of estimation errors where the perturbations are distributed with respect to the standard normal is given by the moment generating function
\begin{align*}
	M(t) &:= \operatorname{det}\bigl(I - 2\Lambda t\bigr)^{-1/2},
\end{align*}
where $\Lambda$ is the diagonal matrix with elements given by the eigenvalues of $\bigl(\DE\bigr)^{\top} \DE$ for any orthonormal basis $\vec{E}(\Pi)$.
\end{prop}
\proof{}
	Let $z$ be a ${d}$-dimensional standard normal Gaussian random variable.
	Then $\Delta = \sum_{k=1}^{d}z_k E_k$ is a perturbation matrix.
	As in Proposition \ref{prop:wce-uniform}, we can write 
	\begin{align*}
					z^\top \bigl(\DE\bigr)^{\top} \DE z 
			=&	z^\top V^\top \Lambda V z, 
	\end{align*}
	where $\Lambda$ is the diagonal matrix of eigenvalues of $\bigl(\DE\bigr)^{\top} \DE$ and $V$ is orthonormal.
	Since $z \sim N(0,I)$ and $VV^\top=I$, we have $w=Vz \sim N(0, VV^\top=I)$. Therefore,
	\begin{align*}
					z^\top V^\top \Lambda V z 
			=&	w^\top \Lambda w 
			=	w^\top \Lambda^{1/2}\Lambda^{1/2} w.
	\end{align*}
	Then $y = \Lambda^{1/2} w \sim N(0, \Lambda)$ and so each component $y_k \sim N(0, \lambda_k)$ and the $y_k$'s are independent.
	Therefore
	\begin{align*}
				w^\top \Lambda^{1/2}\Lambda^{1/2} w 
		=&	y^\top y 
		=	\sum_{k=1}^{d} y_k^2.
	\end{align*}
	The distribution of $y_k^2$ is $\Gamma(1/2,2\lambda_k)$, and thus the sum $\sum_{k=1}^{d}  y_k^2$ has the moment generating function
	\begin{equation*}
		M(t) = \prod_{k=1}^{d} \bigl(1-2\lambda_k t\bigr)^{-1/2},
	\end{equation*}
	where $\lambda_k$ are the eigenvalues of $\bigl(\DE\bigr)^{\top} \DE$.
	 		As in Proposition \ref{prop:worst_case}, the independence of the distribution from the choice of basis $\vec{E}(\Pi_C)$ is a direct result of Proposition \ref{prop:basis_eigenvalue}. 
\endproof
        
\begin{rem}
            A closed form for the density of the distribution found in Proposition \ref{prop:wce-gaussian} is given in equation (7) of \cite{Mathai:SumGamma}.
\end{rem}
				
\begin{figure}
\centering
\includegraphics[width=0.48\textwidth]{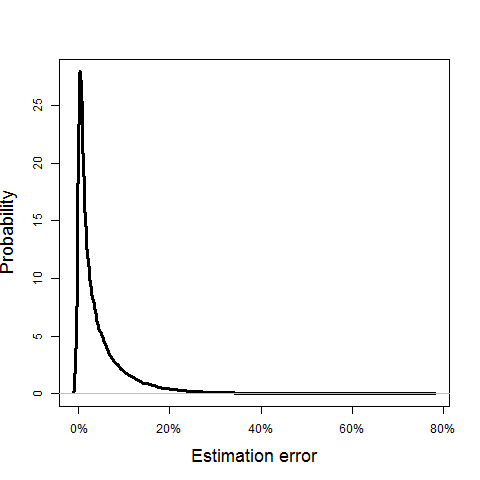}
\hspace{2ex}
\includegraphics[width=0.48\textwidth]{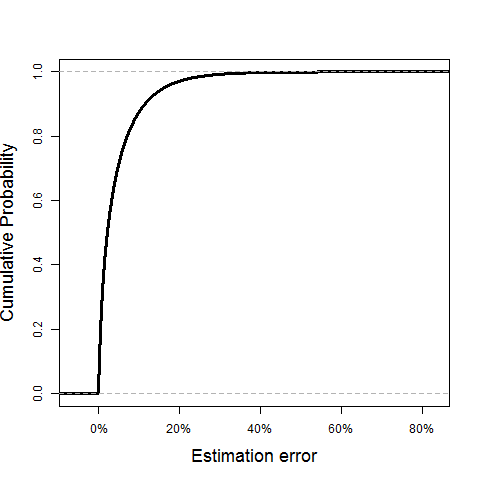}
\caption{The estimated probability density (left) and the CDF (right) of the estimation error, $\| \DB\|_2^2/\|p(\Pi)\|_2^2$, under standard Gaussian perturbations $\Delta$ as described in Example \ref{ex:normal}.}
\label{fig:prob_esterror_gaussian}
\end{figure}

\bexmpl	\label{ex:normal}		
We return again to Example~\ref{ex:clearing_payments} to consider perturbations $\Delta$ sampled from the standard normal distribution.  Figure \ref{fig:prob_esterror_gaussian} shows the density and CDF estimation for the relative estimation error, $\| \DB\|_2^2/\|p(\Pi)\|_2^2$, corresponding to our stylized four-bank network. The probabilities are estimated from 100,000 simulated Gaussian perturbations.
\eexmpl

\subsection{Impact to the payout to society}\label{sec:society}
In this section, we assume that in addition to their interbank liabilities, banks also have a liability to society. Here, society is used as totum pro parte, encompassing all non-financial counterparties, corporate, individual or governmental.
Hence, the set of institutions becomes $\CN_{0} = \{ 0 \} \cup \CN$. Without loss of generality, we assume that all banks $i\in\CN$ owe money to at least one counterparty $j\in\CN_0$ within the system. Otherwise, a bank who owes no money can be absorbed by the society node as it plays the same role within the model structure. The question of interest is then how the payout to society may be mis-estimated (and in particular overestimated) given estimation errors in the relative liabilities matrix. 
This setting has been studied in, e.g., \cite{Glasserman:contagion_review} with the introduction of \textit{outside liabilities}.
We adopt their framework to analyze this question.

The interbank liability matrix $L$ of the previous section is expanded to $L_{0} \in \RR^{(n+1) \times (n+1)}$ given by
\begin{equation*}
L_0 = \left[  
\begin{array}{ccc|c}
0 & \cdots & L_{1n} & L_{10} \\
\vdots & \ddots & \vdots & \vdots\\
L_{n1} & \cdots &0 & L_{n0} \\
\hline 
0 & \cdots & 0 & 0  \\
\end{array}
\right] = \left[
\begin{array}{ccc|c}
~ & ~ & ~ & ~\\
~ & L & ~ & l_0 \\
~ & ~ & ~ & ~\\
\hline
0 & \cdots & 0 & 0\\ 
\end{array}
\right],
\end{equation*}
where $l_0 = \bigl( L_{10}, \cdots, L_{n0} \bigr)^\top$ is the society liability vector. We require that at least one bank has an obligation to society, i.e., $L_{i0}>0$ for some $1\leq i\leq n$.
The total liability of bank $i$ is now given by $\pbar_{i} = \sum_{j = 0}^{n} L_{ij}$.
As stated above, we also require that each bank owes to 
at least one counterparty within the system (possibly society), i.e., $\pbar_i>0$ for all $i \in \CN$.
The relative liability matrix $\Pi_0$ is transformed accordingly,
i.e., $\pi_{ij}\in[0,1]$ and 
$\pi_{ij}=\frac{L_{ij}}{\pbar_i}$. 
An admissible relative liability matrix $\Pi_0$  thus belongs to the set of all right stochastic matrices with entries in $[0,1]$, all diagonal entries 0, and at least one $\pi_{i0}>0$:
\begin{equation*}
	\BA_0 := \lb  \Pi_0 \in [0,1]^{(n+1) \times (n+1)} \; \Big| \; \forall i: \pi_{ii}=0,\; \sum_{j=0}^n \pi_{ij}=1\text{ and } \exists i  \text{ s.t. } \pi_{i0} > 0\rb .
\end{equation*}
An admissible \textit{interbank} relative liability matrix $\Pi$ thus belongs to the set
\begin{equation*}
\BA_{I} := \lb \Pi \in [0,1]^{n \times n} \; \Big| \; \forall i:  \pi_{ii} = 0,\;  \sum_{j=1}^n \pi_{ij} \leq 1 \text{ and }  \exists i \text{ s.t. } \sum_{j=1}^n \pi_{ij} < 1\rb , 
\end{equation*}
which has the same properties as the original interbank relative liability matrix $\BA$ defined in \eqref{def:BA}, except that row sums are smaller or equal to $1$, with at least one strictly smaller than $1$.

The following result is implicitly used in the subsequent sections.  This provides us with the ability to, e.g., consider the directional derivative with respect to the payments made by the $n$ financial firms without considering the societal node (which is equal to $0$ by assumption).
\begin{prop}
If $(\Pi_0 , x , \pbar)$ is a regular network then $I - \text{diag}(d)\Pi$ is invertible.
\end{prop}
\proof{}
This follows immediately from
\[I - \text{diag}(d_0)\Pi_0^\top = \left(\begin{array}{cc} I - \text{diag}(d)\Pi^\top & -\text{diag}(d)\pi_0 \\ \mathbf{0}^\top & 1 \end{array}\right),\]
where $\pi_0 = \bigl(\pi_{10},\cdots,\pi_{n0}\bigr)^\top$ and $d_0$ is the vector of default indicators (of length $n+1$ to include the societal node).
In particular, since $\text{det}(I - \text{diag}(d_0)\Pi_0^\top) \neq 0$ (as shown in the proof of Theorem \ref{thm:derivative}), we can conclude that $\text{det}\bigl(I - \text{diag}(d)\Pi^\top\bigr) \neq 0$.
\endproof

\begin{figure}
\centering
\begin{subfigure}{.45\textwidth}
	\centering
	\includegraphics[width=\linewidth]{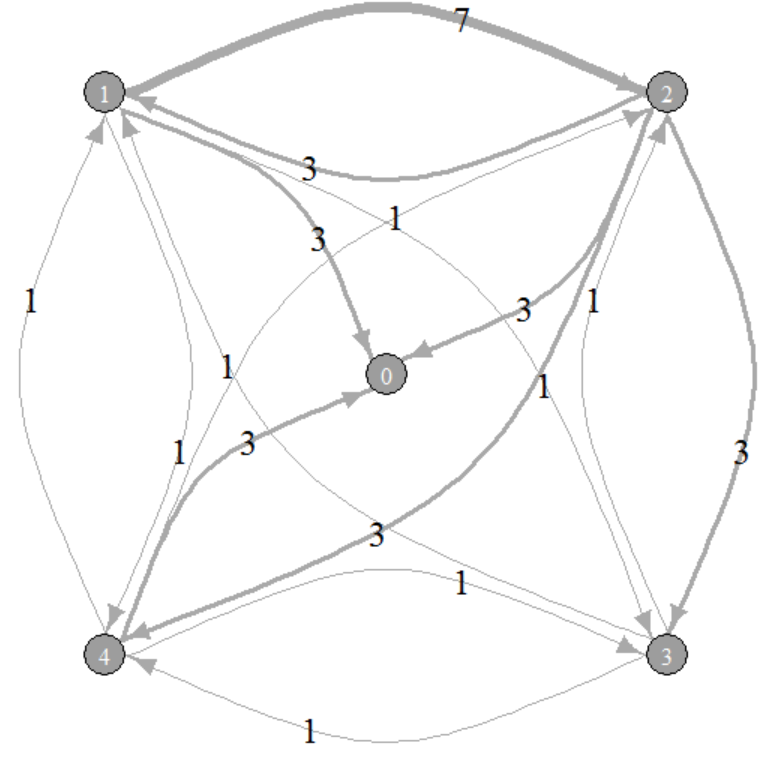}
	\caption{Nominal interbank liabilities}\label{fig:nominal_liabilties_society}
\end{subfigure}
~
\begin{subfigure}{.45\textwidth}
	\centering
	\includegraphics[width=\linewidth]{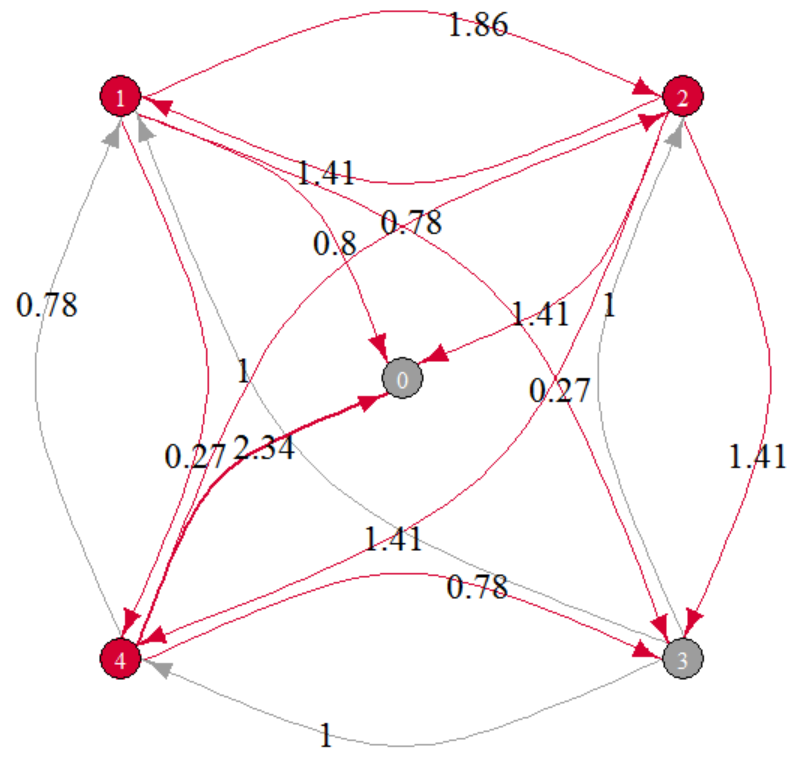}
	\caption{Clearing interbank payments}\label{fig:clearing_payments_society}
\end{subfigure}
\caption{Initial network defined in Example \ref{ex:society}}
\end{figure}

\bexmpl\label{ex:society}
We include now a society node into our example from Section \ref{sec:NetworkModel}. The nominal interbank liabilities and liabilities from each bank to society are shown in Figure \ref{fig:nominal_liabilties_society}. Note that at least one bank has an obligation to society and the society does not owe to any bank. As above, the banks' external assets are given by the vector $x=(0,2,2,2)^\top$. The clearing payments, or the amount of its obligations that each bank is able to repay, is given in Figure \ref{fig:clearing_payments_society}. Banks who are in default are colored red, as are the liabilities that are not repaid in full.
\eexmpl

\subsubsection{Largest reduction in the payout to society}
Next, we use the directional derivative in order to quantify how estimation errors, under $\BB_F^n(\Pi)$ in the interbank relative liability matrix, could lead to an overestimation of the payout to society. 
As it turns out, this problem also has an elegant solution using the basis of perturbation matrices discussed in Appendix \ref{sec:batrices}.
We assume that $(\Pi_{0}, x, \bar{p})$ is a regular financial system and additionally that both the relative liabilities to society $\pi_{0}=(\pi_{10},...,\pi_{n0})^{\top}$ and the total liabilities $\bar{p}$ are exactly known.
\bdfn
Let $(\Pi_{0}, x, \bar{p})$ be a regular financial system. The payout to society is defined as the quantity $\pi_{0}^{\top} p(\Pi)$ where $p(\Pi)$ is the clearing vector of the $n$ firms. 
\edfn
\noindent  Herein we consider the relative liabilities matrix $\Pi_{0}$ to be an estimation of the true relative liabilities. We thus consider the perturbations of the estimated clearing vectors to determine the maximum amount that the payout to society may be overestimated.
To study the optimisation problem of minimizing the payout to society, we assume that at least one bank, but not all banks, default. 
The following proposition shows that this assumption excludes only trivial cases. 
\begin{prop}
Let $(\Pi_{0}, x, \bar{p})$ be a regular system with the interbank relative liability matrix $\Pi \in \BA_{I}$ and $\Delta \in \BB^{n}(\Pi)$. If all banks default, or if no bank defaults, then the payout to society remains unchanged for an arbitrary admissible perturbation $\Delta$. 
\end{prop}
\proof{}
Let $\Delta$ be an arbitrary perturbation matrix. We show that in both cases $\pi_{0}^{\top} \DB = 0$. 
\begin{enumerate}
\item Assume that no bank defaults. Then $\text{diag}(d) = 0$, and the result holds as $\DB = 0$. 
\item Assume all banks default. Then  $\text{diag}(d) = I$.
Hence, $\pi_{0}^{\top} \DB = \pi_{0}^{\top} \left( I - \Pi^{\top} \right)^{-1} \Delta^{\top} p(\Pi)$. 
Note that $\pi_{0}^{\top} \bigl( I - \Pi^{\top} \bigr)^{-1} = \mathbf{1}^{\top}$, because by definition $\pi_{0}^{\top} =  \mathbf{1}^{\top} \bigl( I - \Pi^{\top} \bigr) $. 
Using this and the definitions of $\DB$ and $\Delta$, it follows $ \pi_{0}^{\top} \DB  = \sum_{i = 1}^{n} \sum_{j = 1}^{n} \delta_{ji} p_{j}(\Pi) = 0$.  
\end{enumerate} 
\endproof
\noindent Let $\Delta \in \BB^{n} (\Pi)$  and assume that for a given $h \in \RR: \Pi + h \Delta \in \BA_I$. 
Then, the minimum payout to society is
\begin{equation*}
\min_{\Delta \in \BB^{n} (\Pi)}  \pi_{0}^{\top} p(\Pi + h \Delta). 
\end{equation*}
In order to remove the dependence on $h$ and the magnitude of $\Delta$, we subtract the constant term $\pi_0^{\top} p(\Pi)$ and consider instead
\begin{equation*}
\min_{\Delta \in \BB_{F}^{n} (\Pi )} \lim_{h \to 0} \pi_{0}^{\top}  \frac{ p(\Pi + h \Delta) - p(\Pi)}{ h} = \min_{\Delta \in \BB_{F}^{n} (\Pi )}~  \pi_0^\top \DB.
\end{equation*}

\noindent  As in Section \ref{sec:worst_case}, using the basis of perturbation matrices $\vec{E}(\Pi)$ of $\BB^n(\Pi)$ (see Appendix \ref{sec:batrices}), we can compute the shortfall to society due to perturbations in the relative liability matrix in $\BB_F^n(\Pi)$.
\begin{prop}\label{prop:worst_case_society}
Let $(\Pi_{0}, x , \pbar)$ be a regular financial system. 
The largest shortfall in payments to society due to estimation errors in the liability matrix in $\BB_F^n(\Pi)$ is given by 
\begin{align*}
       \min_{\Delta  \in \BB_{F}^{n} (\Pi )} \pi_0^\top \DB = -\|  
       \pi_{0}^{\top} \DE \|_2.
    \end{align*}
 Furthermore, the largest shortfall to society is achieved by 
 \begin{equation*}
     \Delta_{0}^{*} (\Pi) := -\sum_{k = 1}^{d} \frac{\pi_{0}^{\top} \mathcal{D}_{E_{k}} p (\Pi) }{\| \pi_{0}^{\top} \DE  \|_{2} } E_{k}.
 \end{equation*}
 Additionally, both the largest shortfall and the perturbation matrix that attains that shortfall are independent of the chosen basis $\vec{E}(\Pi)$.
\end{prop}

\proof{}
Since the problem 
\begin{equation*}
\min   \pi_{0}^{\top} \DE  z \; \text{ s.t. } \; \|z\|_2 \leq 1, 
\end{equation*}
has a linear objective, it is equivalent to
\[
\min  \pi_{0}^{\top} \DE z \; \text{ s.t. } \; z^{\top} z = 1 .
\]
By the necessary Karush--Kuhn--Tucker conditions, we know that any solution to this problem must satisfy
\begin{align*}
\bigl(\DE\bigr)^{\top} \pi_{0} + 2\mu z &= 0,\\ z^{\top} z &= 1, 
\end{align*}
for some $\mu \in \mathbb{R}$. The first condition implies $z^* = -\frac{(\DE)^{\top} \pi_{0}}{2\mu}$.
Plugging this into the second implies that $\mu = \pm\frac{\| \pi_{0}^{\top} \DE\|_2}{2} $.  With two possible solutions we plug these back into the original objective to find that the minimum is attained at $\mu = \frac{\| \pi_{0}^{\top} \DE\|_2}{2} $ for an optimal value of:
\begin{equation*}
 \pi_{0}^{\top} \DE z^* = 
 -\frac{ \bigl( \pi_{0}^{\top} \DE \bigr) \bigl( \pi_{0}^{\top} \DE \bigr)^{\top} }{\| \pi_{0}^{\top} \DE \|_2} = 
 - \| \pi_{0}^{\top} \DE\|_2. 
 \end{equation*}
Therefore, the solution is 
\begin{equation*}
    \Delta_{0}^{*} (\Pi) = \sum_{k = 1}^{d} z_k^* E_k = 
-\sum_{k = 1}^{d} \frac{\pi_{0}^{\top} \mathcal{D}_{E_{k}} p (\Pi) }{\| \pi_{0}^{\top} \DE  \|_{2} } E_{k}. 
\end{equation*}
By Proposition \ref{prop:basis_norm}, this result is independent of the choice of basis matrices. 
\endproof

\begin{coro}\label{cor:worst_case_society}
Let $(\Pi_{0}, x , \pbar)$ be a regular financial system. 
The worst case shortfall to society is bounded by
\begin{equation*}
 -\|  \pi_{0}^{\top} \mathcal{D}_{\vec{E}(\Pi_C)}p(\Pi) \|_2
 \leq \min_{\Delta  \in \BBs_{F}^{n} (\Pi )} \pi_0^\top \DB \leq 
-\|\pi_{0}^{\top} \DE \|_2 ,
\end{equation*}
where $\vec{E}(\Pi_C)$ is any orthonormal basis of perturbation matrices of any completely connected network $\Pi_C$. In the case that $\Pi$ itself is a completely connected network then this upper bound is attained.
\end{coro}
\proof{}
This follows by the same logic as Corollary~\ref{cor:worst_case} through the inclusion $\BB_F^n(\Pi) \subseteq \BBs_F^n(\Pi) \subseteq \BB_F^n(\Pi_C)$ for any completely connected network $\Pi_C$.
The independence of this result to the choice of orthonormal basis $\vec{E}(\Pi)$ follows as in Proposition \ref{prop:worst_case_society}. 
\endproof

\begin{figure}
\centering
	\includegraphics[width=.45\linewidth]{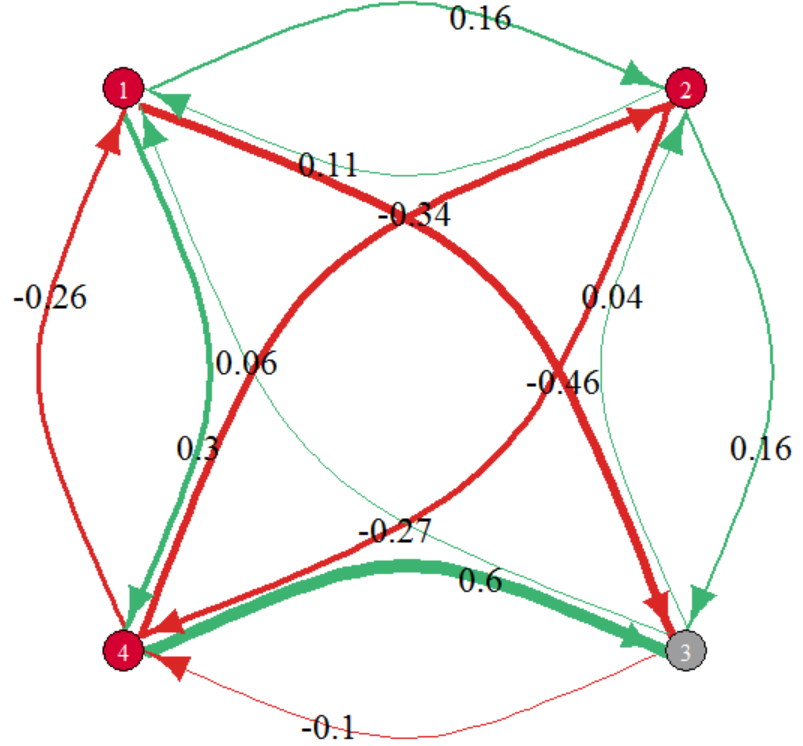}
	\caption{The perturbation in $\BB_F^n(\Pi)$ which generates the largest shortfall for society defined in Example \ref{ex:worstcase_society}.}\label{fig:worstcase_society}
\end{figure}

\bexmpl\label{ex:worstcase_society}
We continue the discussion from Example~\ref{ex:society}: The perturbation resulting in the greatest shortfall for the society's payout, as described in Proposition \ref{prop:worst_case_society}, is given by the matrix
\begin{equation*}
	\Delta_0^*= \left(
	\begin{matrix}
		0 & 0.16 & -0.46 & 0.30 \\
		0.11 & 0 & 0.16 & -0.27 \\
		0.06 & 0.04 & 0 & -0.10 \\
		-0.26 & -0.34 & 0.60 & 0 \\
	\end{matrix}\right).
\end{equation*}
\noindent
This perturbation is depicted in Figure \ref{fig:worstcase_society}. Each edge is labeled with the perturbation of the respective link between banks that achieves this greatest reduction in payout to society. As before, banks who are in default are colored red. The edge linking one node to another is red if the greatest reduction in payout occurs when we have overestimated the value of this link and green if, in the worst case under $\BB_F^n(\Pi)$, we have underestimated the value of this link. 
Edge widths are proportional to the absolute value of the entries in $\Delta_0^*(\Pi)$.
In contrast to Example~\ref{ex:worstcase}, note that $-\Delta_0^*(\Pi)$ is \textbf{not} a solution anymore.
As this network is complete, this also equals the worst case shortfall of $-1.4513$, which is nearly 32\% of the entire estimated payment to society.
 
\eexmpl

\subsubsection{Shortfall to society for uniformly distributed estimation errors}
In this section we compute the reduction in the payout to society when the perturbations are uniformly distributed.
To do so, we consider the linear coefficients $z$ for the basis of perturbation matrices to be chosen uniformly from the ${d}$-dimensional Euclidean unit ball.  
Then $\Delta = \sum_{k = 1}^{d}z_k E_k$ is a perturbation matrix.

\begin{prop}
Let $(\Pi_{0}, x , \pbar)$ be a regular financial system. 
The distribution of changes in payments to society where the perturbations are uniformly distributed on the unit ball is given by 
\begin{align*}
    \mathds{P}\bigl(\pi_{0}^{\top} \DB \leq \alpha\bigr)	&=	
								\frac{1}{2}
									+\frac{\alpha}{\bigl\|\bigl(\DE\bigr)^{\top} \pi_0\bigr\|_2}
										\frac{\Gamma(1+\frac{{d}}{2})}{\sqrt{\pi}\Gamma(\frac{1+{d}}{2})}
										\,_2F_1\Biggl(\frac{1}{2},\frac{1-{d}}{2};\frac{3}{2};
										\frac{\alpha^2}
											{\bigl\|\bigl(\DE\bigr)^{\top} \pi_0\bigr\|_2^2}\Biggr)
\end{align*}
for $\alpha \in [-\bigl\|\bigl(\DE\bigr)^{\top} \pi_0\bigr\|_2,\bigl\|\bigl(\DE\bigr)^{\top} \pi_0\bigr\|_2]$ and 0 for $\alpha \leq -\bigl\|\bigl(\DE\bigr)^{\top} \pi_0\bigr\|_2$ and 1 for $ \alpha \geq \|\bigl(\DE\bigr)^{\top} \pi_0\|_2 $.  In the above equation, $_2F_1$ is the standard hypergeometric function.  Furthermore, this distribution holds for any choice of basis matrices $\vec{E}(\Pi)$.
\end{prop}
\proof{}
	Let $z$ be a uniform random variable on the unit ball in $\RR^{d}$ centered at the origin. 
	Then $\Delta = \sum_{k=1}^{d}z_k E_k$ is a perturbation matrix.
	Note that by linearity of the directional derivative, we have
	\begin{equation*}
		\mathcal{D}_\Delta\bigl(\pi_0^\top p(\Pi)\bigr)
		= \pi_0^\top \DE z,
	\end{equation*}
	where 
	$\DE 
		= \bigl(\mathcal{D}_{E_1}\bigl(p(\Pi)\bigr), \, \ldots \,, \mathcal{D}_{E_{{d}}}\bigl(p(\Pi)\bigr)\bigr).$
	Since $z$ is uniform on the unit ball,
\begin{align}
\nonumber &	\mathds{P}\Bigl(\mathcal{D}_\Delta\bigl(\pi_0^\top p(\Pi)\bigr) \leq \alpha\Bigr) 
=	\mathds{P}\bigl(\pi_0^\top \DE z \leq \alpha\bigr) \\
\label{eq:uniform1} 
&=	\frac{\text{vol}\bigl( \bigl\{ z \in \RR^{d} \; \big|\; \pi_0^\top \DE z \leq \alpha , z^\top z \leq 1 \bigr\} \bigr)}
{\text{vol}( \{ z \in \RR^{d} \;|\; z^\top z \leq 1 \} )} 
\\ \nonumber 
&=	\frac{\text{vol}\Bigl( \Bigl\{ z \in \RR^{d} \;\Big|\; \bigl(\bigl(\DE\bigr)^{\top} \pi_0\bigr)^\top z \leq \alpha ,\; z^\top z \leq 1 \Bigr\} \Bigr)}
{\text{vol}( \{ z \in \RR^{d} \;|\; z^\top z \leq 1 \} )} \\
\nonumber 
&=	\frac{\text{vol}\biggl( \biggl\{ z \in \RR^{d}\;\bigg|\; 
\biggl(\frac{(\DE)^{\top} \pi_0}
		{\|(\DE)^{\top} \pi_0\|_2}\biggr)^\top z 
	\leq 
		\frac{\alpha}{\|(\DE)^{\top} \pi_0\|_2} , 
		z^\top z \leq 1  \biggr\} \biggr)}
{\text{vol}( \{ z \in \RR^{d} \;|\; z^\top z \leq 1 \} )} \\
\label{eq:uniform2} 
&=	\frac{\text{vol}\biggl( \biggl\{ z \in \RR^{d}\;\bigg|\; e_1^\top z 
	\leq 
		\frac{\alpha}
				{\|(\DE)^{\top} \pi_0\|_2} ,\; 
		z^\top z \leq 1 \biggr\} \biggr)}
{\text{vol}( \{ z \in \RR^{d} \;|\; z^\top z \leq 1 \} )} 
\\
\nonumber &= 
\begin{cases}
0 &\text{if } \alpha < -\bigl\|\bigl(\DE\bigr)^{\top} \pi_0\bigr\|_2\\
\frac{1}{2} I_{\theta}(\frac{1+d}{2},\frac{1}{2}) &\text{if }  \alpha \in \bigl\|\bigl(\DE\bigr)^{\top} \pi_0\bigr\|_2 \times [-1,0]\\
1 - \frac{1}{2} I_{\theta}(\frac{1+d}{2},\frac{1}{2}) & \text{if }  \alpha \in \bigl\|\bigl(\DE\bigr)^{\top} \pi_0\bigr\|_2 \times [0,1]\\
1 &\text{if } \alpha > \bigl\|\bigl(\DE\bigr)^{\top} \pi_0\bigr\|_2
\end{cases}, 
\qquad \theta = \frac{\bigl\|\bigl(\DE\bigr)^{\top} \pi_0\bigr\|_2^2 - \alpha^2}{\bigl\|\bigl(\DE\bigr)^{\top} \pi_0\bigr\|_2^2}
\\
\nonumber &= 	
\begin{cases}
0 &\text{if } \alpha < -\bigl\|\bigl(\DE\bigr)^{\top} \pi_0\bigr\|_2\\ 
\frac{1}{2}
	+\frac{\alpha}{\|(\DE)^{\top} \pi_0\|_2}
		\frac{\Gamma(1+\frac{{d}}{2})}{\sqrt{\pi}\Gamma(\frac{1+{d}}{2})}
		\,_2F_1\biggl(\frac{1}{2},\frac{1-{d}}{2};\frac{3}{2};
		\frac{\alpha^2}
			{\|(\DE)^{\top} \pi_0\|_2^2}\biggr)
	&\text{if } \alpha \in \bigl\|\bigl(\DE\bigr)^{\top} \pi_0\bigr\|_2\times[-1,1]\\ 
	1 &\text{if } \alpha > \bigl\|\bigl(\DE\bigr)^{\top} \pi_0\bigr\|_2\end{cases},
	\end{align}
	where $I_{\theta}(a,b)$ is the regularized incomplete beta function (see, e.g., \cite[Chapter 8.17]{NIST:DLMF}) and $_2F_1$ is the standard hypergeometric function (see, e.g., \cite[Chapter 15]{NIST:DLMF}).  Equation~\eqref{eq:uniform1} follows from considering the probability by taking the ratio of the volume of the fraction of the unit ball satisfying the probability event to the full volume of the unit ball.  Equation~\eqref{eq:uniform2} follows by symmetry of the unit ball and since $\bigl(\DE\bigr)^{\top} \pi_0/\|\bigl(\DE\bigr)^{\top} \pi_0\|_2$ has unit norm.  The penultimate result follows directly from the volume of the spherical cap (see, e.g., \cite[Equation (2)]{li2011concise}). 
The final result follows from properties of the regularized incomplete beta function (see, e.g., \cite[Chapter 8.17]{NIST:DLMF}), i.e.,
	\[I_{\theta}\biggl(\frac{1+d}{2},\frac{1}{2}\biggr) = 1 - 2\sqrt{1-\theta}\frac{\Gamma(1+\frac{d}{2})}{\sqrt{\pi}\Gamma(\frac{1+d}{2})} \, _2F_1\biggl(\frac{1}{2},\frac{1-{d}}{2};\frac{3}{2};1-\theta\biggr), \] 
with $\theta = \frac{\bigl\|\bigl(\DE\bigr)^{\top} \pi_0\bigr\|_2^2 - \alpha^2}{\bigl\|\bigl(\DE\bigr)^{\top} \pi_0\bigr\|_2^2}$,	and noting that the case for $\alpha$ positive and negative can be written under the same equation using the standard hypergeometric function.
	The independence of this result to the choice of orthonormal basis $\vec{E}(\Pi)$ follows as in Proposition \ref{prop:worst_case_society} as the distribution only depends on the basis $\vec{E}(\Pi)$ through the norm $\|\bigl(\DE\bigr)^{\top} \pi_0\|_2$. 
\endproof

\begin{figure}
\centering
\includegraphics[width=0.48\textwidth]{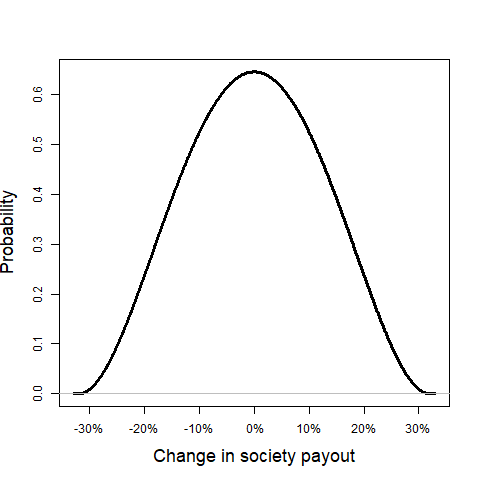}
\hspace{2ex}
\includegraphics[width=0.48\textwidth]{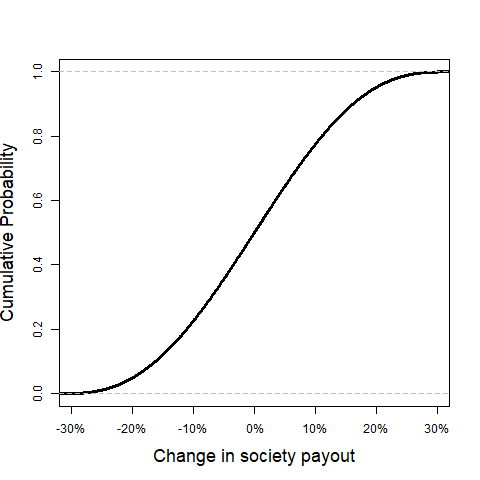}
\caption{Estimated probability density (left) and CDF (right) of relative reduction in payout to society, $\frac{\pi_0^\top\DB}{\pi_0^\top p(\Pi)}$, under uniform perturbations $\Delta$ as described in Example \ref{ex:uniform-society}}
\label{fig:prob_socpay_ball}
\end{figure}

\bexmpl	\label{ex:uniform-society}		
We return to Example~\ref{ex:society} and consider perturbations $\Delta$ sampled from the uniform distribution. 
The left and right panels of Figure \ref{fig:prob_socpay_ball} show the density and the CDF respectively for the relative reduction in society payout under uniformly distributed errors in our stylized four-bank network. Figure \ref{fig:CI_socpay_ball} shows both the largest reduction and increase in the payout to society as well as various confidence intervals for the change in the payout as a function of the perturbation size, $h$.  As $h^{*}$ and $h^{**}$ depend on the choice of perturbation matrix $\Delta$, we present the confidence intervals on an extrapolated interval for $h \in [0,1]$.
\eexmpl

\subsubsection{Shortfall to society for normally distributed estimation errors}
We will now consider the same problem as above under the assumptions that the errors follow a standard normal distribution.  As in Section~\ref{sec:worst_case_gaussian}, we note that the magnitude of the perturbations is no longer bounded by 1 in this setting.

\begin{prop}
Let $(\Pi_{0}, x , \pbar)$ be a regular financial system. 
The distribution of changes to payments to society where the perturbations follow a multivariate standard normal distribution is given by
\begin{align*}
				\mathcal{D}_\Delta\bigl(\pi_0^\top p(\Pi)\bigr) &\sim 
				N\Bigl(0, \bigl\|\bigl(\DE\bigr)^{\top} \pi_0\bigr\|_2^2\Bigr).
\end{align*}
Furthermore, this distribution holds for any choice of basis matrices $\vec{E}(\Pi)$.
\end{prop}
\proof{}
Let $z$ be a ${d}$-dimensional standard normal Gaussian random variable. 
The result follows immediately by linearity and affine transformations of the multivariate Gaussian distribution.
The independence of this result to the choice of orthonormal basis $\vec{E}(\Pi)$ follows as in Proposition \ref{prop:worst_case_society} as the distribution only depends on the basis $\vec{E}(\Pi)$ through the norm $\bigl\|\bigl(\DE\bigr)^{\top} \pi_0\bigr\|_2$.
\endproof

\begin{figure}
\centering
\begin{subfigure}{.48\textwidth}
	\centering
	\includegraphics[width=\linewidth]{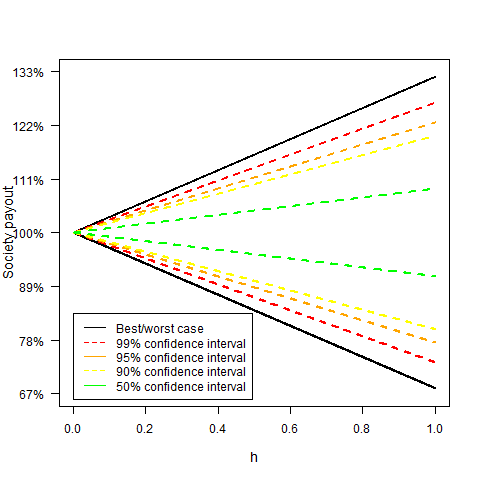}
	\caption{Uniformly distributed perturbations}\label{fig:CI_socpay_ball}
\end{subfigure}
~
\begin{subfigure}{.48\textwidth}
	\centering
	\includegraphics[width=\linewidth]{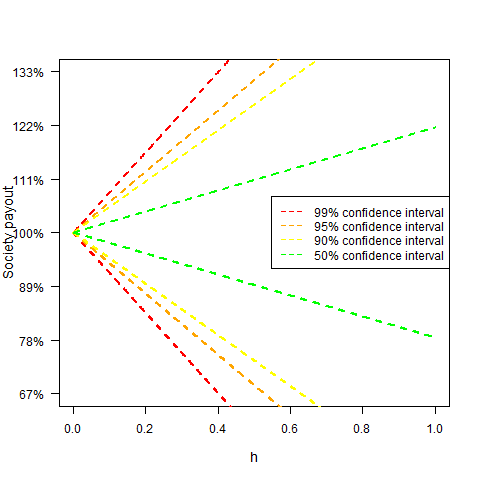}
	\caption{Normally distributed perturbations}\label{fig:CI_socpay_gaussian}
\end{subfigure}
\caption{Largest increase and decrease of the payout to society and confidence intervals for the payout as a function of the size $h$ of perturbations in $\BB_F^n(\Pi)$, respectively $\BB^n(\Pi)$, where perturbations  $\Delta$ are sampled uniformly (left) and from the standard Gaussian distribution (right) for the stylized four-bank system with society as described in Example \ref{ex:society}. 
}
\end{figure}

\bexmpl \label{ex:normal-society}			
We return once more to Example~\ref{ex:society} to consider perturbations $\Delta$ sampled from the standard normal distribution.  Figure \ref{fig:CI_socpay_gaussian} shows various confidence intervals for the relative change in payout to society under normally distributed errors under $\BB^4(\Pi)$, as a function of the perturbation size, $h$.  As $h^{*}$ and $h^{**}$ depend on the choice of perturbation matrix $\Delta$ we present the confidence intervals on an extrapolated interval for $h \in [0,1]$.
\eexmpl

\section{Empirical application: assessing the robustness of systemic risk analyses} \label{sec:EBA}
In this section, we study the robustness of conclusions that can be drawn from systemic risk studies that use the Eisenberg--Noe algorithm to model direct contagion.
We use the same dataset from 2011 of European banks from the European Banking Authority that has been used in previous studies relying on the Eisenberg--Noe framework (\cite{gandy_bayesian_2015,ChenLiuYao:ModelingFinancialSystemicRisk_NetworkEffectMarketLiquidityEffect2014}).  
As in these papers, given the heuristic approach to the dataset, our exercise should be considered to be an illustration of our results and methodology, rather than a realistic full-fledged empirical analysis. 

With respect to the model's data requirements, the EBA dataset only provides information on the total assets $TA_{i}$, the capital $c_{i}$ and a proxy for interbank exposures, $a_{i}^{IB}$. 
To populate the remaining key variables of the Eisenberg--Noe model, we therefore first assume, as in \cite{ChenLiuYao:ModelingFinancialSystemicRisk_NetworkEffectMarketLiquidityEffect2014}, that for each bank the interbank liabilities are equal to the interbank assets. 
Furthermore, we assume that all non-interbank assets are external assets, and the non-interbank liabilities are liabilities to a society sink-node. 
Hence,
\begin{align*}
l_{i}^{IB} & := a_{i}^{IB}, \\
L_{i0} & := TA_i - l_{i}^{IB}  - c_{i}, \\
a_{i}^{0} &  := TA_{i}  - a_{i}^{IB}.
\end{align*}
Consequently, the Eisenberg--Noe model variables are
\begin{align*}
& \text{Total liabilities: } \bar{p}_{i}  = L_{i0}  + l_{i}^{IB}, \\
& \text{Total external assets: } x_{i}  =   a_{i}^{0}.  
\end{align*}
\noindent
Note that each bank's net worth hence exactly corresponds to the book value of equity, or the banks' capitals: $TA_{i} - \bar{p}_{i} = a_{i}^{0} + a_{i}^{IB}  - l_{i}^{IB} - L_{i0} = c_{i}$ .   

The final key ingredient to the model is the (relative) liabilities matrix. 
This is usually highly confidential data, and is not provided in the EBA data set. 
In \cite{gandy_bayesian_2015}, Gandy and Veraart propose an elegant Bayesian sampling methodology to generate individual interbank liabilities, given information on the total interbank liabilities and total interbank assets of each bank. 
The authors have developed an $R$-package called ``systemicrisk" that implements a Gibbs sampler to generate samples from this conditional distribution. 
As our analysis requires an initial liability matrix, we use the European Banking Authority (EBA) data as input to their code in order to generate such a liability matrix.
As suggested by \cite[Section 5.3]{gandy_bayesian_2015}, we perturb the interbank liabilities $l_{i}^{IB}$ slightly (such that they are not exactly equal to the interbank assets, while keeping the total sums equal) to fulfill the condition that $L$ be connected along rows and columns. 
We then run their algorithm, with parameters $p = 0.5, \text{thinning} = 10^{4}, n_{burn-in} = 10^{9}, \lambda = \frac{p n (n-1)}{\sum_{i = 1}^{N} a_{i}^{IB}}  \approx 1.2178 10^{-3}$,
to create one realisation of a 87 $\times$ 87 network of banks from the data.
(We needed to exclude banks DE029, LU45 and SI058 because the mapping of the data to the model as described above created violations of the conditions for the algorithm and resulted in an error message.)

\begin{figure}
\centering
\begin{subfigure}{.48\textwidth}
	\centering
	\includegraphics[width=\linewidth]{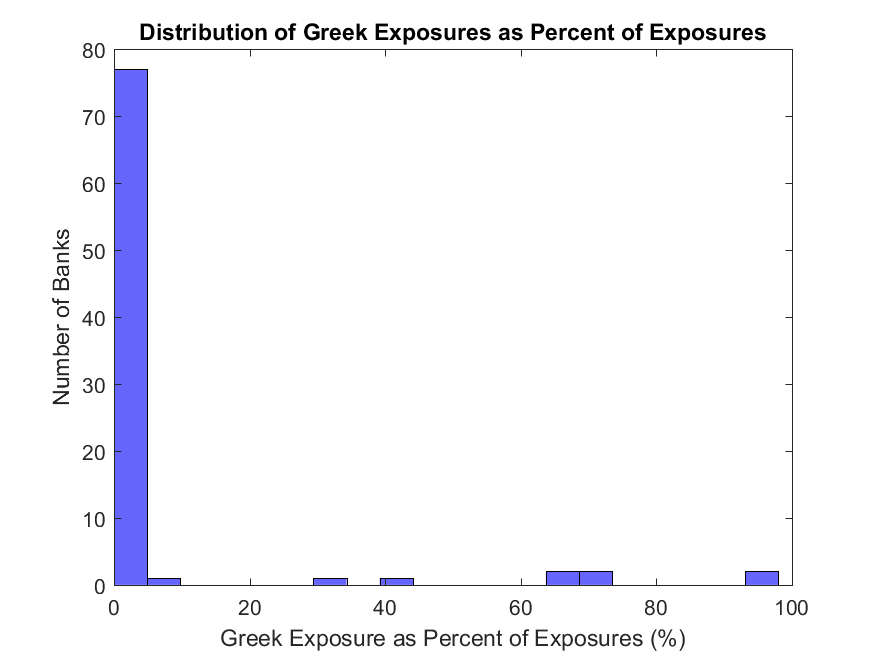}
	\caption{Distribution of Greek exposures}\label{fig:EBA_GreekExposures}
\end{subfigure}
~
\begin{subfigure}{.48\textwidth}
	\centering
	\includegraphics[width=\linewidth]{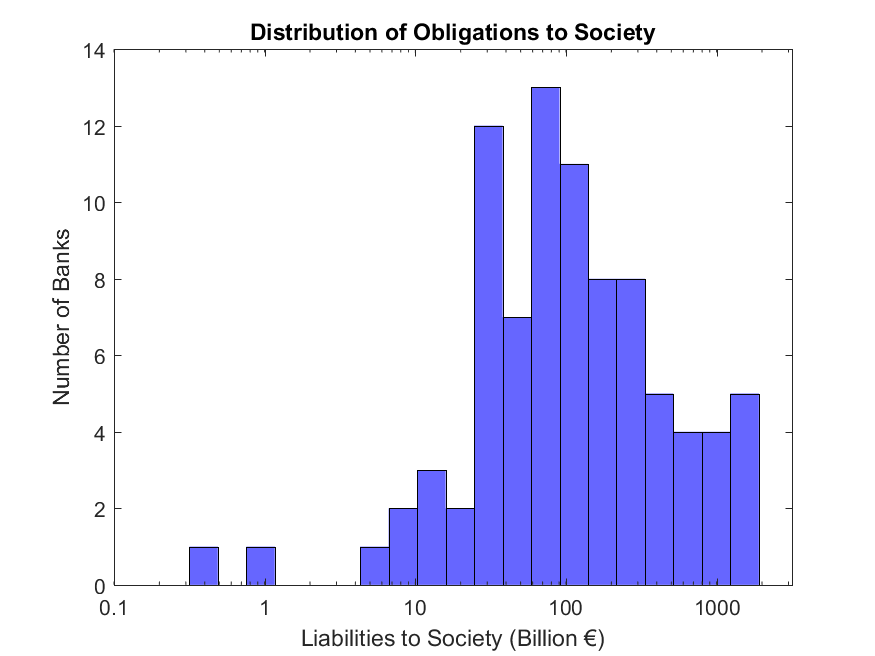}
	\caption{Distribution of bank obligations to society}\label{fig:EBA_L0-log}
\end{subfigure}
\caption{Histograms of data from the EBA dataset. }
\end{figure}

For simplicity and to consider an extreme event that would trigger a systemic crisis in the European banking system, we analyze what might have happened if Greece had defaulted on its debt and exited the Eurozone. We study this shock by decreasing the external assets of each bank by its individual Greek exposures, i.e. setting Greek bond values to zero.  {The histogram of Greek exposures (as a percentage of total exposures), displayed in Figure \ref{fig:EBA_GreekExposures}, shows a large heterogeneity of exposures, with the majority of banks having no (or negligible) exposures to Greece, but a small number of Greek banks having substantial exposures to Greece (between 64\% - 96\% of total assets).}
In our sensitivity analysis we resample the underlying liabilities matrix from the Gandy \& Veraart algorithm \cite{gandy_bayesian_2015} 1000 times.

In each of our 1000 simulated networks considered there were 9 specific institutions that default on their debts in the Eisenberg--Noe framework; in only 3 simulated networks (0.3\% of all simulations) there were between 1 and 3 additional banks that fail.  As such, the traditional analysis of sensitivity of the Eisenberg--Noe framework would conclude that this contagion model is robust to errors in the relative liabilities matrix.  This is consistent with the work of, e.g., \cite{GlassermanYoung:ContagionFinancialNetworks2014}.

\begin{figure}
\centering
\begin{subfigure}{.48\textwidth}
	\centering
	\includegraphics[width=\linewidth]{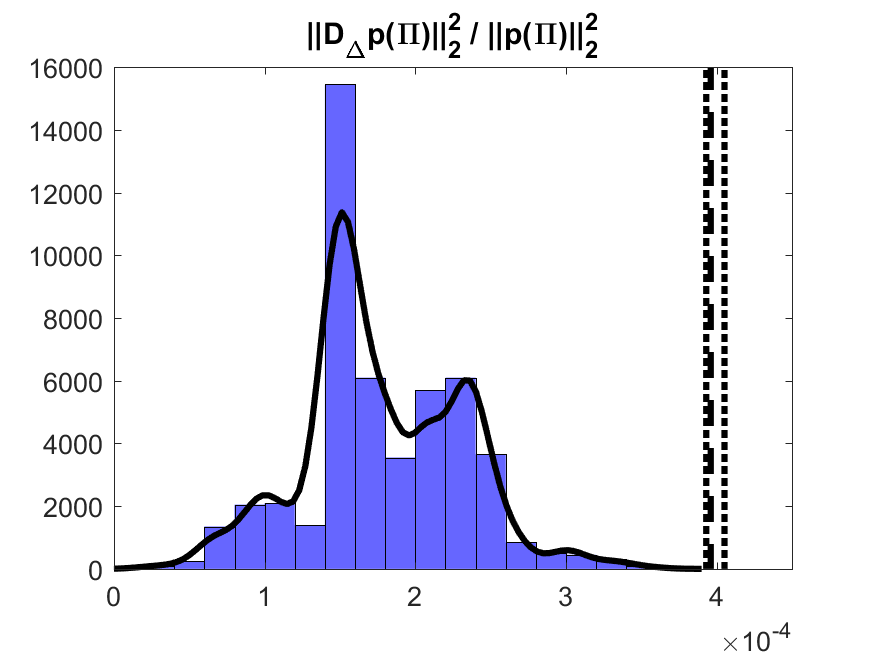}
	\caption{Relative error of the clearing vector.}\label{fig:EBA_worst_estimation_error}
\end{subfigure}
~
\begin{subfigure}{.48\textwidth}
	\centering
	\includegraphics[width=\linewidth]{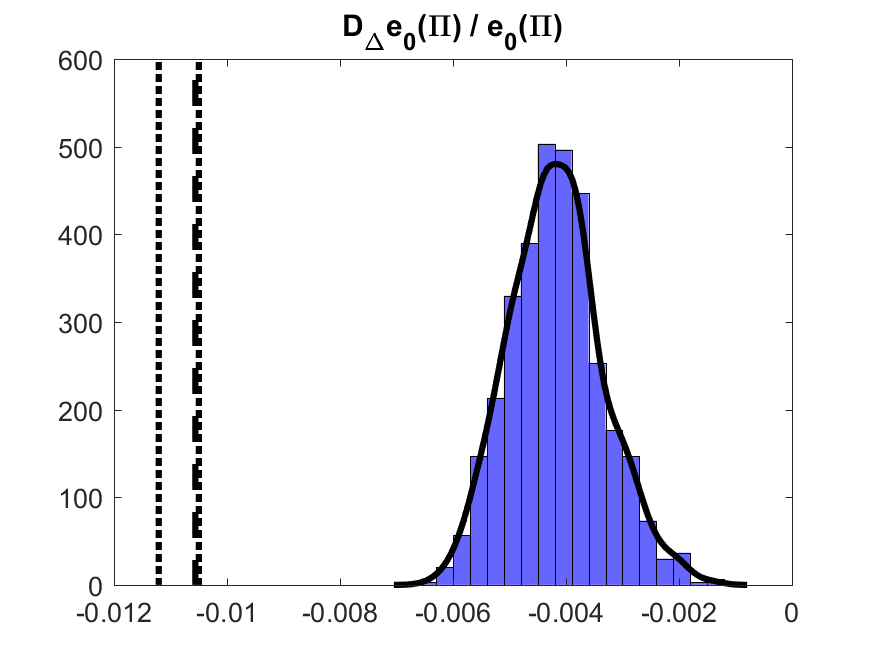}
	\caption{Relative error of the payments to society.}\label{fig:EBA_society}
\end{subfigure}\\
\begin{subfigure}{.48\textwidth}
	\centering
	\includegraphics[width=\linewidth]{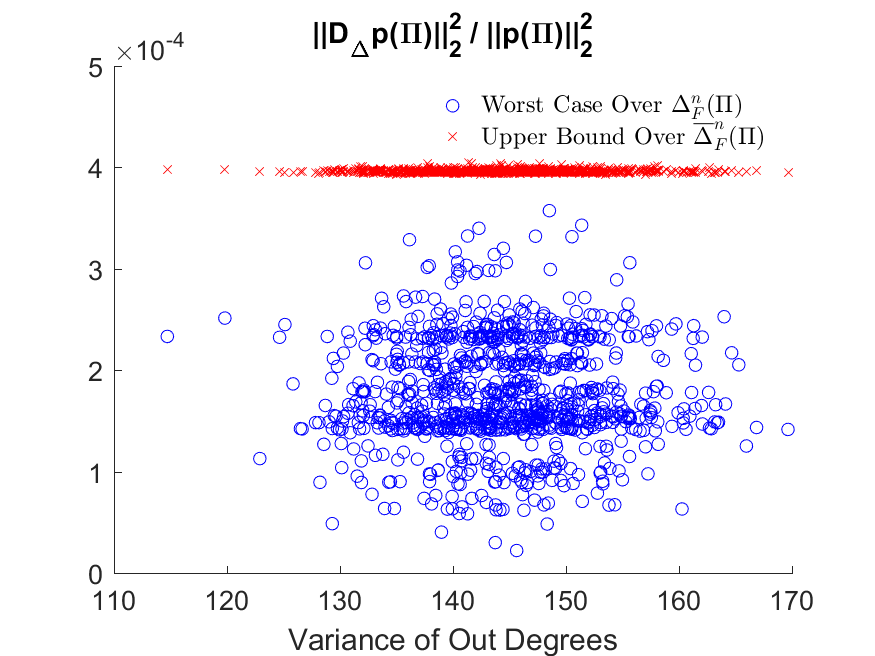}
	\caption{{There is no clear dependence between relative error of the clearing vector and network heterogeneity.}}\label{fig:EBA_var_worst_estimation_error}
\end{subfigure}
~
\begin{subfigure}{.48\textwidth}
	\centering
	\includegraphics[width=\linewidth]{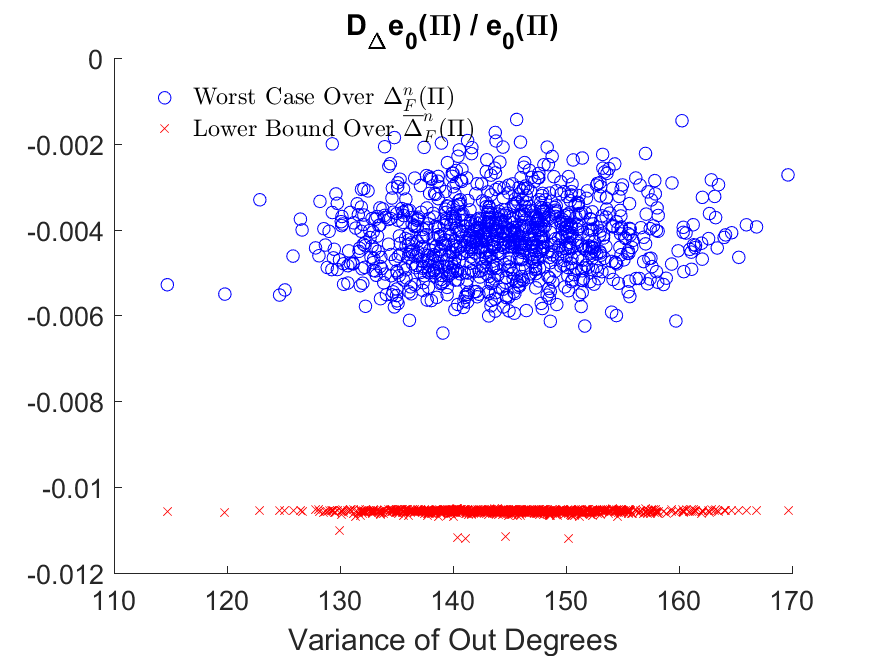}
	\caption{{There is no clear relationship between the relative error of the payments to society and network heterogeneity}.}\label{fig:EBA_var_society}
\end{subfigure}
\caption{Top: Empirical densities of the relative errors in the Eisenberg--Noe framework as a function of random networks calibrated to the same EBA dataset. The dotted vertical lines indicate the maximal and minimal empirical values of the upper bound of the worst case and the dashed line indicates the median upper bound. 
{Bottom: Dependence of the clearing vector perturbation on network heterogeneity. }}
\end{figure}

However, we now consider the maximal deviation in both the estimation errors and the payments to society in each of our 1000 simulated networks under $\BB_F^n(\Pi)$. 
{The societal obligations are the same in all 1000 simulated networks, and their histogram, depicted in Figure \ref{fig:EBA_L0-log}, reveals as for the Greek exposures, considerable heterogeneity.}
Figure \ref{fig:EBA_worst_estimation_error} depicts the empirical density of the maximal deviation estimation errors $\frac{\| D_\Delta p(\Pi) \|_2^2}{\| p(\Pi) \|_2^2}$ for $\Delta \in \BB_F^n(\Pi)$.  Figure \ref{fig:EBA_society} depicts the empirical density of maximal fractional shortfalls to society $\frac{ D_\Delta e_0(\Pi) }{e_0(\Pi)}$.  We also depict the upper bound of the worst case perturbation errors for each of the 1000 simulated networks.

Notably in Figure \ref{fig:EBA_worst_estimation_error} we see that the shape of the network, calibrated to the same EBA data set, can vastly change the impact that the worst case estimation error has under perturbations in $\BB_F^n(\Pi)$.  In this plot of the empirical densities, we see the range of normalized worst case first order estimation errors range from 0 to nearly $4 \times 10^{-4}$.  That is a 0 to 2\% normed deviation of the clearing payments (while the value of $\|p(\Pi)\|_2$ itself has only minor variations: a total range of under 27 million EUR compared to its norm of near 5 trillion EUR for the different simulated networks $\Pi$).  
The upper bound on these perturbation errors (for the norm rather than norm squared) is approximately 2\%, and as can be seen in Figure \ref{fig:EBA_worst_estimation_error}, the range of obtained upper bounds is very small.  This indicates that such a bound is rather insensitive to the initial relative liability matrix $\Pi$.  Therefore any such computed upper bound is of value to a regulator, even if the initial estimate of the relative liabilities $\Pi$ is incorrect.

When we consider instead Figure \ref{fig:EBA_society} we see that the density is more bell shaped, again with a large variation from the least change (roughly $-0.001$) to the most change (roughly $-0.007$) in the normalized impact to society; this proves as with Figure \ref{fig:EBA_worst_estimation_error} that the underlying network can provide large differences in the apparent stability of a simulation to validation.  
While these values may appear small, the $10^{-3}$ arises from normalising the deviation of the clearing vector with the value of the societal node but still amounts to a variation on the order of 23.2 - 162.4 billion EUR.
Thus this sensitivity is as if entire banks' assets vanished from the wealth of society.
The upper bound of these perturbation errors is approximately twice as high as the obtained maximal deviations computed under $\BB^n_F(\Pi)$.  Notably, the median upper bound of the worst case error is nearly equal to the minimum possible value, though with a skinny tail reaching off to greater errors.

{Finally, Figures \ref{fig:EBA_var_worst_estimation_error} and \ref{fig:EBA_var_society} analyze the impact of network heterogeneity on the perturbation of the clearing vector. 
To this end, we quantify ``network heterogeneity" as the variance of the degree distribution of out-edges. 
It varies between 110 to 170 in the 1000 simulated networks, thus displaying a reasonable level of heterogeneity. 
Figure \ref{fig:EBA_var_worst_estimation_error} shows the worst case relative error over $\BB^n_F(\Pi)$ (blue circles) and $\BBs^{n}_{F}(\Pi)$ (red crosses) respectively.
Similarly, Figure \ref{fig:EBA_var_society} shows a scatter plot of the relative error of the payment to society against the variance of the degree distribution in the network. 
Neither figure seems to suggest a clear relation between the relative errors and the network heterogeneity. 
Note that Figures \ref{fig:EBA_worst_estimation_error} and \ref{fig:EBA_society} are obtained by projecting all points onto the $y$-axis in Figures \ref{fig:EBA_var_worst_estimation_error} and \ref{fig:EBA_var_society}. 
}

\section{Conclusion}\label{sec:conc}
In this paper we analyse the sensitivity of clearing payments {in the standard Eisenberg--Noe framework} to misspecification or estimation errors in the relative liabilities matrix. 
We accomplish this by determining the directional derivative of the clearing payments with respect to the relative liabilities matrix.
We extend this result to consider the full Taylor expansion of the fixed points to determine the clearing payments as a closed-form perturbation of an initial solution.

We further study worst case and probabilistic interpretations of our perturbation analysis. 
 {In this simple setting,} our results provide an upper bound on the largest shift for the clearing vector {as well as} a lower bound for the shortfall to society. 
In a numerical case study of the European banking system, we demonstrate that, even when the set of defaulting firms remains constant, the clearing payments and wealth of society can be greatly impacted.  This is true even in the case that the existence and non-existence of links is pre-specified. When the existence and non-existence of links is unknown, then the upper bound of the errors can be utilized which generally provides errors that are significantly less sensitive to the initial estimate of the relative liabilities and roughly twice as large as the errors under pre-specification of links.

{Our sensitivity analysis is based on the standard Eisenberg--Noe model. As such, it omits a number of other important extensions that have been developed in the literature (such as bankruptcy costs, fire sales, or the impact of the network topology). 
For a full quantification of risk and uncertainty, future research will therefore need to develop a model that combines -- and weighs -- all of these relevant channels of contagion.  
Nevertheless, our results provide a first step towards quantifying the impact of estimation errors in the interbank liability matrix and thereby improving tools for systemic risk analysis.}

\section{Acknowledgements}
The collaboration leading to this article was initiated at the AMS Mathematics Research Community 2015 in Financial Mathematics, supported by the National Science Foundation, Division of Mathematical Sciences, under Grant No.\ 1321794.
Eric Schaanning's PhD studies were kindly funded by the Fonds National de la Recherche Luxembourg under its AFR PhD grant scheme. 
\bibliographystyle{informs2014} 
\bibliography{references_Sensitivity_paper} 

\appendix
\section{Appendix}
\subsection{Proofs}
\paragraph{Proof of Proposition \ref{prop:continuity}}

\proof{}
This proof follows the logic of \cite[Lemma 5.2]{feinstein_measures_2015} and  \cite[Theorem 4]{RenYuanJiang:Framework}.
Fix the net assets $x$ and total obligation $\pbar$. Let $\phi:[0,\pbar]\times \BA \rightarrow [0,\bar{p}]$ be the function defined by $\phi(\hat{p},\Pi) := \bigl( \phi_1(\hat{p},\Pi),\cdots,\phi_n(\hat{p},\Pi) \bigr)^\top$, where
\begin{equation*}
\phi_i(\hat{p},\Pi)=\pbar_i \wedge \biggl(x_i + \sum_{j=1}^n \pi_{ji} \hat{p}_j\biggr),\quad i\in \CN.
\end{equation*}
The function $\phi$ is jointly continuous with respect to the payment vector $\hat{p}$ and the relative liabilities $\pi_{ij}$ for $i,j \in\CN$. 
Because the system is regular and thus has a unique fixed point, it follows from \cite[Proposition A.2]{feinstein_measures_2015} that the graph 
\[
\text{graph}(p) = \bigl\{(\Pi,\hat{p}) \in \BA\times [0,\pbar] \; \big| \; \phi(\hat{p},\Pi) = \hat{p}\bigr\}
\]
 is closed.
		Define the projection 
			$\Psi:  \BA \times [0,\pbar] \rightarrow \BA$ 
		as
			$\Psi(\Pi,p)=\Pi$.
By \cite[Proposition A.3]{feinstein_measures_2015}, $\Psi$ is a closed mapping in the product topology.
Then, in order to show that $p$ is continuous, take $U\subset [0,\pbar]$ closed. Then
\begin{equation*}
p^{-1}[ U ]	
=\bigl\lbrace \Pi \in \BA \; \big| \; p(\Pi)\in U \bigr\rbrace
=\Psi\bigl(\text{graph}(p)\cap(\BA\times U)\bigr).
\end{equation*}
The graph of $p$ is closed and  $\BA$ is closed by definition.
Hence $p^{-1}[U]$ is closed and the function $p$ is continuous with respect to $\Pi$.  
\endproof

\paragraph{Proof of Theorem \ref{thm:derivative}}
We note that our proof does not assume \textit{a priori} that the clearing vector $p$ is differentiable; we comment on this simpler case below.
  
\proof{}
We assume that the net external assets lie in the set
\[\biggl\lbrace x \in \RR_{+}^{n} \, \Big\vert \, \nexists i \in \CN \text{ s.t. } x_{i} + \sum_{j = 1}^n \pi_{ji}p_j(\Pi) = \pbar_{i} \biggr\rbrace.\]
Denote  $\alpha^{(1)} = x + \Pi^{\top} p(\Pi) = (\alpha^{(1)}_1,\cdots,\alpha^{(1)}_n)^\top$ and $\alpha^{(2)} =  x + ( \Pi + h \Delta )^{\top} p ( \Pi + h \Delta ) = (\alpha^{(2)}_1,\cdots,\alpha^{(2)}_n)^\top$. 
By continuity of $p$ with respect to $\Pi$ (Proposition \ref{prop:continuity}) we have for all $ i\in\CN$, $\alpha^{(2)}_i \rightarrow \alpha^{(1)}_i$  as $h \rightarrow 0$ and thus
 $\one_{\{ \{ \alpha^{(1)}_i < \pbar_i\} \cap \{ \alpha^{(2)}_i > \pbar_i\} \} }  \to 0$,  $\one_{ \{ \{\alpha^{(1)}_i >  \pbar_i \} \cap \{ \alpha^{(2)}_i < \pbar_i \}\} } \to 0$ and  $\one_{ \{\{ \alpha^{(1)}_i < \pbar_i \} \cap \{ \alpha^{(2)}_i < \pbar_i \}\} } \rightarrow \one_{\{ \alpha^{(1)}_i < \pbar_i \}}$. 
To prove the existence of $\DB$, we will show that the following two limits,
\begin{equation*}
\begin{split}
\overline{\DB}_i &= \limsup_{h\to 0 } \frac{p_i(\Pi+h \Delta) - p_i(\Pi)}{h} \qquad \text{ and}\\
\underline{\DB}_i &= \liminf_{h\to 0} \frac{p_i(\Pi+h \Delta) - p_i(\Pi)}{h}\\
\end{split}
\end{equation*}
are equal for each component. Consider the upper limit

\begin{equation*}
\begin{split}
& \overline{\DB}_i  = \limsup_{h \to 0}  \frac{p_i(\Pi+h \Delta)  - p_i(\Pi ) }{ h }  \\
&	= \limsup_{h \to 0} \frac{1}{h} \Biggl( \biggl( \pbar_i \wedge ( x_i + \sum_{j=1}^n(\pi_{ji}+h \delta_{ji}) p_j(\Pi + h \Delta) ) \biggr) - 
	\biggl( \pbar_i \wedge ( x_i + \sum_{j=1}^n \pi_{ji} p_j ( \Pi ) )\biggr) \Biggr) \\
& = \limsup_{h \to 0} \Biggl( 
	0 \times \one_{ \{ \{  \alpha^{(1)}_i > \pbar_i \} \cap  \{\alpha^{(2)}_i > \pbar_i \}\} } 
	+ \frac{\pbar_i - (x_i + \sum_{j=1}^n \pi_{ji} p_j(\Pi))}{h} \one_{ \{\{ \alpha^{(1)}_i < \pbar_i \} \cap \{ \alpha^{(2)}_i > \pbar_i \}\} } \Biggr. \\
	& \quad + \frac{ x_i + \sum_{j=1}^n \pi_{ji} p_j(\Pi + h \Delta) + h \sum_{j=1}^n \delta_{ji} p_j(\Pi + h \Delta) - \pbar_i}{h} 
	\one_{ \{\{ \alpha^{(1)}_i >  \pbar_i \} \cap \{ \alpha^{(2)}_i < \pbar_i \} \}} \quad  \\
	& \quad + \Biggl.  \frac{ \sum_{j=1}^n \pi_{ji} \bigl( p_j(\Pi + h \Delta) - p_j(\Pi) \bigr) + h \sum_{j=1}^n \delta_{ji} p_j( \Pi + h \Delta ) }{h} 
	\one_{ \{\{ \alpha^{(1)}_i < \pbar_i \} \cap \{ \alpha^{(2)}_i < \pbar_i \} \}} 	
	\Biggr) \\
	& = \biggl(  \sum_{j=1}^n \pi_{ji} \overline{\DB}_j +  \sum_{j=1}^n \delta_{ji} p_j(\Pi) \biggr) \one_{ \{ \alpha^{(1)}_i < \pbar_i \} }   \\
& = d_i \sum_{j=1}^n \pi_{ji} \overline{\DB}_j + d_i \sum_{j=1}^n \delta_{ji} p_j(\Pi) =: \Psi_i\bigl(\overline{\DB}\bigr)
\end{split}
\end{equation*}
for some function $\Psi: \RR^n \to \RR^n$.

Similarly, we get 
\begin{equation*}
\begin{split}
\underline{\DB}_i & =d_i \sum_{j=1}^n \pi_{ji} \underline{\DB}_j + d_i \sum_{j=1}^n \delta_{ji} p_j(\Pi) = \Psi_i\bigl(\underline{\DB}\bigr).\\
\end{split}
\end{equation*}
Hence, both $\overline{\DB}$ and $\underline{\DB}$ are fixed points of the same mapping $\Psi$. Assuming that this fixed point problem has a unique solution it follows
$$
\overline{\DB}_i = \underline{\DB}_i,
$$
for all $i \in \CN$.
Therefore, under this assumption, $\DB$ is well defined and it is the solution to the fixed point equation
\begin{equation*}
\DB = \Psi\bigl(\DB\bigr) = \text{diag}(d) \Pi^{\top} \DB + \text{diag}(d) \Delta^{\top} p(\Pi). 
\end{equation*}
Next, we proceed to show that $\bigl( I - \text{diag}(d) \Pi^{\top} \bigr)$ is invertible, which establishes uniqueness of the fixed point and the directional derivative \eqref{eqn:directional derivative} to conclude the proof.

First, assume that $\text{diag}(d) \Pi^\top$ is irreducible, i.e., the graph with adjacency matrix $\text{diag}(d) \Pi^\top$ has directed paths in both directions between any two vertices $i \neq j$. Then by the Perron--Frobenius Theorem (see, e.g., \cite[Section 8.7.2]{gentle_matrix_2007}), $\text{diag}(d) \Pi^\top$ has an eigenvector $v>\mathbf{0}$  corresponding to eigenvalue $\rho(\text{diag}(d) \Pi^\top)$, where $\rho(\cdot)$ is the spectral radius of a matrix.  As eigenvectors are only unique up to a multiplicative constant, we may assume $\|v\|_1=1$. Under the assumption of a regular system, at least one bank must be solvent, i.e., there exists some $i$ such that $\text{diag}(d)_{ii}=0$. This implies that there exists a column such that the column sum of $\text{diag}(d) \Pi^\top$ is strictly less than 1. In fact, any insolvent institution $j$ with obligations to bank $i$ will have column sum of $\text{diag}(d) \Pi^\top$ strictly less than 1. If all banks are solvent, $\text{diag}(d)$ is the zero matrix and the result is trivial.
Thus there is some matrix $M\geq0$, $M\neq0$ so that each column sum of $\text{diag}(d) \Pi^\top + M$ is 1, i.e.
\[
	\mathbf{1}^\top \bigl(\text{diag}(d) \Pi^\top+M\bigr)=\mathbf{1}^\top.
\]
Note that the column sums of $\text{diag}(d) \Pi^\top$ are at most 1 since each row sum of $\Pi$ is 1. Therefore the spectral radius of $\text{diag}(d) \Pi^\top$ must be less than or equal to $1$. Moreover, we must have $\rho(\text{diag}(d) \Pi^\top)<1$. Otherwise, $\rho(\text{diag}(d) \Pi^\top)=1$, which along with the scaling of the eigenvector so that $\|v\|_1 = 1$ implies
\begin{equation*}
\begin{aligned}
		1
	=	\mathbf{1}^\top v
	=	\mathbf{1}^\top \bigl(\text{diag}(d) \Pi^\top+M\bigr) v
	=	\mathbf{1}^\top (v+Mv) 
	=	1 + \mathbf{1}^\top Mv
	>	1 ,
\end{aligned}
\end{equation*}
as $\Pi^\top v = 1v$ by the definition of eigenvalues. Therefore, we can conclude that, in the case $\text{diag}(d) \Pi^\top$ is irreducible, $\rho(\text{diag}(d) \Pi^\top)<1$.

Now suppose that $\text{diag}(d) \Pi^\top$ is reducible, i.e., $\text{diag}(d) \Pi^\top$ is similar to a block upper triangular matrix $D$, with irreducible diagonal blocks $D_i$, $i=1,\ldots,m$ for some $m<n$. Under the assumption of a regular system, each $D_i$ has at least one column whose sum is strictly less than 1. As in the preceding case, this implies that $\rho(D_i)<1$ for each $i$ and therefore
\begin{equation*}
	\rho(\text{diag}(d) \Pi^\top) =	\rho(D) <	1.
\end{equation*}
Since the maximal eigenvalue of $\text{diag}(d) \Pi^\top$ is strictly less than 1, 0 cannot be an eigenvalue of $I-\text{diag}(d) \Pi^\top$. This suffices to show that $I-\text{diag}(d) \Pi^\top$ is invertible. 
\endproof

\begin{rem}
If one assumes that $p$ is differentiable with respect to the relative liabilities $\Pi$, the result of Theorem~\ref{thm:derivative} can be obtained directly from implicit differentiation of the representation \[p(\Pi) = (I - \text{diag}(d))\bar p + \text{diag}(d)[x + \Pi^\top p(\Pi)].\]
\end{rem}

\paragraph{Proof of Theorem \ref{thm:higher_order_derivatives}}

\proof{}
We prove the result by induction. Theorem \ref{thm:derivative} shows the result for $k = 1$. 
We now assume that equation \eqref{eq:nderivative} holds for $k$ and we proceed to show that it holds for $k+1$.
As in Theorem \ref{thm:derivative}, we show the existence of \eqref{eqn:higher order derivative} by computing the two limits:
\begin{equation*}
\begin{split}
\overline{\mathcal{D}_{\Delta}^{(k+1)} p(\Pi)}_i &= \limsup_{h\to 0} \frac{\mathcal{D}_{\Delta}^{(k)} p(\Pi+h\Delta)_i  - \mathcal{D}_{\Delta}^{(k)} p(\Pi)_i }{h} \qquad \text{ and}\\
\underline{\mathcal{D}_{\Delta}^{(k+1)} p(\Pi)}_i &= \liminf_{h\to 0} \frac{\mathcal{D}_{\Delta}^{(k)} p(\Pi+h\Delta)_i  - \mathcal{D}_{\Delta}^{(k)} p(\Pi)_i }{h}. 
\end{split}
\end{equation*}
The first order Taylor approximation for matrix inverses gives by the differentiation rules for the matrix inverse (cf. \cite[p. 152]{gentle_matrix_2007}) for $X,Y \in \RR^{n \times n}$ and $h$ small enough: $(X + h Y )^{-1} \approx X^{-1} - h X^{-1} Y X^{-1}$.
Applying this fact with $X=I-\text{diag}(d)\Pi^T$ and $Y=-\text{diag}(d)\Delta^T$, we have
\begin{equation*}
\bigl(I - \text{diag}(d) (\Pi + h \Delta)^{\top} \bigr)^{-1} \approx \bigl(I - \text{diag}(d) \Pi^{\top} \bigr)^{-1} + h \bigl(I - \text{diag}(d) \Pi^{\top} \bigr)^{-1} \text{diag}(d) \Delta^{\top} \bigl(I - \text{diag}(d) \Pi^{\top} \bigr)^{-1}.
\end{equation*}
Additionally, we note that the $k^{th}$ order derivative, similarly to all lower order derivatives, is continuous with respect to the relative liabilities matrix $\Pi$ since (by assumption of the induction) $\mathcal{D}_{\Delta}^{(k)} p(\Pi) = k! \bigl( \bigl( I - \text{diag}(d) \Pi^{\top} \bigr)^{-1} \text{diag}(d) \Delta^{\top} \bigr)^{k} p(\Pi)$, where $p(\Pi)$ and $\bigl(  I - \text{diag}(d) \Pi^{\top}  \bigr)^{-1}$ are both continuous with respect to $\Pi$ (see Proposition \ref{prop:continuity} and the continuity of the matrix inverse).
Consider now the upper limit

\begin{equation*}
\begin{aligned}
& \overline{\mathcal{D}_{\Delta}^{(k+1)} p(\Pi)} = \limsup_{h\to 0} \frac{\mathcal{D}_{\Delta}^{(k)} p(\Pi+h\Delta)  - \mathcal{D}_{\Delta}^{(k)} p(\Pi) }{h}\\
&= \limsup_{h \to 0} \frac{k}{h}\Bigl(\bigl(I - \text{diag}(d)(\Pi + h\Delta)^\top\bigr)^{-1}\text{diag}(d)\Delta^\top \mathcal{D}_{\Delta}^{(k-1)}p(\Pi + h\Delta)\Bigr.\\
& \qquad\qquad\qquad \Bigl. - \bigl(I - \text{diag}(d)\Pi^\top\bigl)^{-1} \text{diag}(d)\Delta^\top \mathcal{D}_{\Delta}^{(k-1)}p(\Pi)\Bigr)\\
&= \limsup_{h\to 0} k\bigl(I-\text{diag}(d) \Pi^\top\bigr)^{-1} \text{diag}(d) \Delta^\top\frac{\mathcal{D}_{\Delta}^{(k-1)} p(\Pi+h\Delta)-\mathcal{D}_{\Delta}^{(k-1)} p(\Pi)}{h} \\
& \quad + \limsup_{h\to 0} \frac{k\,h}{h}\bigl(I-\text{diag}(d) \Pi^\top\bigr)^{-1} \text{diag}(d) \Delta^\top \bigl(I-\text{diag}(d) \Pi^\top\bigr)^{-1} \text{diag}(d) \Delta^{\top} \mathcal{D}_{\Delta}^{(k-1)} p(\Pi+h\Delta)\\
&= k\bigl(I-\text{diag}(d) \Pi^\top\bigr)^{-1} \text{diag}(d) \Delta^\top \mathcal{D}_{\Delta}^{(k)} p(\Pi) \\
& \quad + k\bigl(I-\text{diag}(d) \Pi^\top\bigr)^{-1} \text{diag}(d) \Delta^\top \bigl(I-\text{diag}(d) \Pi^\top\bigr)^{-1} \text{diag}(d) \Delta^{\top} \mathcal{D}_{\Delta}^{(k-1)} p(\Pi)\\
&=k\bigl(I-\text{diag}(d) \Pi^\top\bigr)^{-1} \text{diag}(d) \Delta^\top \mathcal{D}_{\Delta}^{(k)} p(\Pi) + \bigl(I-\text{diag}(d) \Pi^\top\bigr)^{-1} \text{diag}(d) \Delta^{\top} \mathcal{D}_{\Delta}^{(k)} p(\Pi)\\
&=(k+1)\bigl(I-\text{diag}(d) \Pi^\top\bigr)^{-1} \text{diag}(d) \Delta^\top \mathcal{D}_{\Delta}^{(k)} p(\Pi). 
\end{aligned}
\end{equation*}
Similarly, we obtain $\underline{\mathcal{D}_{\Delta}^{(k+1)} p(\Pi)} = (k+1)\bigl(I-\text{diag}(d) \Pi^\top\bigr)^{-1} \text{diag}(d) \Delta^\top \mathcal{D}_{\Delta}^{(k)} p(\Pi)$. The existence of the limit and the result \eqref{eq:nderivative} follow for all $k \geq 1$. 

With the above results on all $k^{th}$ order directional derivatives, we now consider the full Taylor expansion.
First, by the definition of $h^{**}$ given in \eqref{eqn:hstarstar}, $\text{diag}(d)$ is fixed for $h \in (-h^{**},h^{**})$. 
By the definition of the clearing payments $p$ (given in \eqref{def:pBar}) and defaulting firms $\text{diag}(d)$ (defined in Theorem \ref{thm:derivative}), along with the fact that $I-\text{diag}(d)(\Pi+h\Delta)^\top$ is invertible (as shown in the proof of Theorem \ref{thm:derivative} since $(\Pi + h\Delta,x,\pbar)$ remains a regular system by $h \in (-h^{**},h^{**}) \subseteq (-h^*,h^*)$), we have
\begin{equation}\label{eq:prepresentation}
\begin{aligned}
p(\Pi+h\Delta)
&=\text{diag}(d)\bigl(x+(\Pi+h\Delta)^\top p(\Pi+h\Delta)\bigr)+\bigl(I-\text{diag}(d)\bigr)\bar{p}\\
&=\bigl(I-\text{diag}(d)(\Pi+h\Delta)^\top\bigr)^{-1}\Bigl(\text{diag}(d) x+\bigl(I-\text{diag}(d)\bigr)\bar{p}\Bigr).
\end{aligned}
\end{equation}
Similarly we find that
\begin{equation}\label{eq:prepresentation2}
p(\Pi) = \bigl(I-\text{diag}(d)\Pi^\top\bigr)^{-1}\Bigl(\text{diag}(d) x+ \bigl(I-\text{diag}(d)\bigr)\bar{p}\Bigr).
\end{equation}
By combining \eqref{eq:prepresentation} and \eqref{eq:prepresentation2}, we immediately find 
\begin{equation*}
p(\Pi + h\Delta) = \bigl(I - \text{diag}(d)(\Pi + h \Delta)^{\top}\bigr)^{-1} \bigl(I - \text{diag}(d)\Pi^{\top}\bigr)p(\Pi).
\end{equation*}
Additionally, we can show that 
\[
\bigl(I - \text{diag}(d)(\Pi + h \Delta)^{\top}\bigr)^{-1} \bigl(I - \text{diag}(d)\Pi^{\top}\bigr) = \Bigl(I-h\bigl(I-\text{diag}(d)\Pi^\top\bigr)^{-1}\text{diag}(d) \Delta^\top\Bigr)^{-1}
\]
directly by 
\begin{align*}
\bigl(I - \text{diag}(d) &  (\Pi + h \Delta)^{\top}\bigr)^{-1} \bigl(I - \text{diag}(d)\Pi^{\top}\bigr) \Bigl(I-h\bigl(I-\text{diag}(d)\Pi^\top\bigr)^{-1}\text{diag}(d) \Delta^\top\Bigr)\\
&= \bigl(I - \text{diag}(d)(\Pi + h \Delta)^{\top}\bigr)^{-1} \bigl(I - \text{diag}(d)\Pi^{\top} - h\,\text{diag}(d) \Delta^{\top}\bigr)\\
&= \bigl(I - \text{diag}(d)(\Pi + h \Delta)^{\top}\bigr)^{-1} \bigl(I - \text{diag}(d)(\Pi + h\Delta)^\top\bigr) = I.
\end{align*}
Therefore, for any $h \in (-h^{**},h^{**})$, we find
\[p(\Pi + h\Delta) = \Bigl(I-h\bigl(I-\text{diag}(d)\Pi^\top\bigr)^{-1}\text{diag}(d) \Delta^\top\Bigr)^{-1} p(\Pi),\]
i.e., \eqref{eq:taylor}.

Now let us consider the perturbations of size $h$ within the neighbourhood  
\[\mathcal{H} := \left\lbrace h \in \RR \; \Bigg| \;  
    |h| < \min\Biggl\{h^{**}, \frac{1}{\rho\Bigl(\bigl( I - \text{diag}(d) \Pi^{\top} \bigr)^{-1} \text{diag}(d) \Delta^{\top}\Bigr)}\Biggr\}\\ 
\right\rbrace  . \]
We will employ the following property of matrix inverses (see \cite[p. 126]{meyer_matrix_2000}): If $X, Y \in \RR^{n \times n}$ so that $X^{-1}$ exists and $\lim_{k\rightarrow\infty}(X^{-1}Y)^k=0$, then
\begin{equation*}
(X+Y)^{-1}=\sum_{k=0}^\infty\bigl(-X^{-1}Y\bigr)^k X^{-1}.
\end{equation*}
We take $X=I-\text{diag}(d)\Pi^\top$ and $Y=-h\,\text{diag}(d)\Delta^\top$.
Since $\rho\bigl(h\bigl( I - \text{diag}(d) \Pi^{\top} \bigr)^{-1} \text{diag}(d) \Delta^{\top}\bigr) = |h| \rho\bigl(\bigl( I - \text{diag}(d) \Pi^{\top} \bigr)^{-1} \text{diag}(d) \Delta^{\top}\bigr)<1$ by the assumption that $|h| <\frac{1}{\rho\bigl(\bigl( I - \text{diag}(d) \Pi^{\top} \bigr)^{-1} \text{diag}(d) \Delta^{\top}\bigr)}$, we have
\begin{equation*}
\lim_{k\rightarrow\infty}\Bigl[h\bigl(I-\text{diag}(d)\Pi^\top\bigr)^{-1}\text{diag}(d) \Delta^\top\Bigr]^k=0 ,
\end{equation*}
using a property of the spectral radius (see \cite[p.\ 617]{meyer_matrix_2000}). 
Thus, by combining this result with~\eqref{eq:prepresentation}, we have
\begin{equation*}
\begin{aligned}
& p(\Pi+h\Delta)
=\bigl(I-\text{diag}(d)(\Pi+h\Delta)^\top\bigr)^{-1}\Bigl(\text{diag}(d) x+\bigl(I-\text{diag}(d)\bigr)\bar{p}\Bigr) \\
&=\sum_{k=0}^\infty\Bigl(h\bigl(I-\text{diag}(d)\Pi^\top\bigr)^{-1}\text{diag}(d) \Delta^\top\Bigr)^k \bigl(I-\text{diag}(d)\Pi^\top\bigr)^{-1} \Bigl(\text{diag}(d) x+\bigl(I-\text{diag}(d)\bigr)\bar{p}\Bigr) \\
&=\sum_{k=0}^\infty\Bigl(h\bigl(I-\text{diag}(d)\Pi^\top\bigr)^{-1}\text{diag}(d) \Delta^\top\Bigr)^k p(\Pi) \\
&= \sum_{k = 0}^\infty \frac{h^k}{k!} \mathcal{D}_{\Delta}^{(k)} p(\Pi).
\end{aligned}
\end{equation*}
The penultimate equality above follows directly from~\eqref{eq:prepresentation2}.   The last equality follows directly from the definition of the $k^{th}$ order directional derivatives proven above.  Thus we have shown the full Taylor expansion is exact on $\mathcal{H} \subseteq (-h^{**},h^{**})$.  

Finally, since we have already shown that \eqref{eq:taylor} is exact for any $h \in (-h^{**},h^{**})$ and 
\[
\Bigl( - h \bigl( I - \text{diag}(d) \Pi^{\top} \bigr)^{-1} \text{diag}(d) \Delta^{\top} \Bigr)
\]
is singular for at least one of the elements $h \in \bigl\{-\frac{1}{\rho(( I - \text{diag}(d) \Pi^{\top} )^{-1} \text{diag}(d) \Delta^{\top})},\frac{1}{\rho(( I - \text{diag}(d) \Pi^{\top} )^{-1} \text{diag}(d) \Delta^{\top})}\bigr\}$ by construction, it must follow that $h^{**} \leq \frac{1}{\rho(( I - \text{diag}(d) \Pi^{\top} )^{-1} \text{diag}(d) \Delta^{\top})}$.  That is, $\mathcal{H} = (-h^{**},h^{**})$.

\endproof

\subsection{An orthonormal basis for perturbation matrices}\label{sec:batrices}
We construct here an orthonormal basis for the matrices in $\BB^n(\Pi)$. 
To fix ideas, consider the case $n=4$, where the general form of a matrix $\Delta \in \BB^{4}(\Pi_C)$ for a fully connected network $\Pi_{C}$ can be written as
\begin{equation*}
\BB^4 (\Pi_{C} ) =\left\{
\text{diag} (\pbar)^{-1} \left.
\left(
\begin{array}{cccc} 
0 & z_1 & z_2  & -z_1-z_2 \\ 
z_3 & 0 & z_4 & -z_{3} - z_{4} \\ 
z_5 & - \sum_{k = 1}^{5} z_{k} & 0 & \sum_{k = 1}^{4} z_{k} \\ 
-z_{3} - z_{5} & \sum_{k = 2}^{5} z_{k} & - z_{2} - z_{4} & 0 
\end{array}
\right) \quad
\right| \, \,
z \in \RR^{5}
\right\},
\end{equation*}
from which it is clear that there are $5$ degrees of freedom. 
It is easy to see that in general one has $d = n^2 - 3n + 1$ degrees of freedom. 
In the case $n=4$, two such 
basis elements $\hat E_1$ and $\hat E_2$ are given by
\begin{equation*}
\hat E_1=
\left(
\begin{array}{cccc}
0 & \frac{1}{\pbar_{1}} & 0 & \frac{-1}{\pbar_{1}} \\ 
0 & 0 & 0 & 0 \\
0 & \frac{-1}{\pbar_{3}} & 0 & \frac{1}{\pbar_{3}} \\
0 & 0 & 0 & 0 
\end{array}
\right) \qquad \text{ and } 
\qquad
\hat E_2=
\left(
\begin{array}{cccc}
0 & 0 & \frac{1}{\pbar_{1}}  & \frac{-1}{\pbar_{1}} \\ 
0 & 0 & 0 & 0 \\
0 & \frac{-1}{\pbar_{3}} & 0 & \frac{1}{\pbar_{3}} \\
0 & \frac{1}{\pbar_{4}}  & \frac{-1}{\pbar_{4}} & 0 
\end{array}
\right). 
\end{equation*}

\noindent In general we note that $\BB^n(\Pi)$ is a closed, convex polyhedral set; we will take advantage of this fact in order to generate a general method for constructing basis matrices for $\BB^n(\Pi)$, as follows:
\begin{enumerate}
\item Define 
\begin{align*}
\vec{\BB}^{n} (\Pi) := \bigg\{  \delta \in \RR^{n^{2}} \; \bigg| \;  & 
\delta_{i +n(i-1)} = 0, \quad \sum_{j = 1}^{n} \delta_{i + n(j-1)} = 0, \\
&  \sum_{j = 1}^{n} \bar{p}_{j} \delta_{n(i-1) + j} = 0, \quad \one_{\{\pi_{ij} = 0\}}\delta_{i + n(j-1)} = 0 \; \forall i,j \bigg\}  
\end{align*}    
to be a vectorised version of $\BB^n(\Pi) $.

\item Construct a matrix $A(\Pi) \in \RR^{(n^2 + 2n)\times n^2}$ so that $\vec{\BB}^n(\Pi) = \{\delta \in \RR^{n^2} \; | \; A(\Pi)\delta = 0\}$. Note that the total degrees of freedom for $\vec{\BB}^n(\Pi)$ (and therefore also for $\BB^n(\Pi)$) is given by the rank of the matrix $A(\Pi)$. We include enough rows in the matrix $A(\Pi)$ in order to ensure that the $n$ row sums and $n$ (weighted) column sums are 0 and that components of $\delta$ are equal to zero based on $\pi_{ij} = 0$.
\item  An orthonormal basis of $\vec{\BB}^n(\Pi)$ can be found by generating the orthonormal basis $\{e_1,...,e_{d}\}$ of the null space of $A(\Pi)$.
\item Finally our basis matrices $\{E_1,...,E_{d}\}$ can be generated by reshaping the basis of the null space of $A(\Pi)$ by setting $E_{k;i,j} := e_{k;i+n(j-1)}$ for any $k = 1,...,d$ and $i,j \in \CN$.
\end{enumerate}

\bdfn
The set 
\begin{equation*}
\vec{E}^n (\Pi) := \lb E_{1}, \ldots, E_{d} \rb 
\end{equation*}
is an \textbf{orthonormal basis of perturbation matrices}  for the relative liability matrix $\Pi$. 
Additionally, the vector
\begin{equation*}
\DE := \bigl( \mathcal{D}_{E_{1}} p(\Pi) , \ldots,  \mathcal{D}_{E_{d}} p(\Pi)  \bigr) \in \RR^{n \times d}
\end{equation*}
is a \textbf{vector of basis directional derivatives} for the relative liability matrix $\Pi$.
\edfn

\noindent 
We define two matrices to be orthogonal when their vectorised forms are orthogonal in $\RR^{n^{2}}$, and note that, by construction, any matrix in the basis of perturbation matrices $\vec{E}^n(\Pi)$ has unit Frobenius norm.


\begin{prop}\label{prop:basis_eigenvalue}
Let $\Pi \in \BA$. Then the set of eigenvalues of $\bigl(\DE\bigr)^{\top} \DE$ is the same for any choice of orthonormal basis of perturbation matrices $\vec{E}(\Pi)$.  Additionally, if $z\bigl(\lambda,\vec{E}(\Pi)\bigr) \in \RR^{d}$ is the eigenvector corresponding to eigenvalue $\lambda$ and basis $\vec{E}(\Pi)$, then $\sum_{k = 1}^{d} z_k\bigl(\lambda,\vec{E}(\Pi)\bigr) E_k$ is independent of the choice of basis.
\end{prop}
\proof{}
Let $E$ be the vectorised version of $\vec{E}(\Pi)$ and let $F \neq E$ be a different orthonormal basis.  By linearity of the directional derivative (see Theorem \ref{thm:derivative}) we can immediately state that $\bigl(\DE\bigr)^{\top} \DE = E^\top C E$ for some matrix $C \in \RR^{n^2 \times n^2}$.  Let $(\lambda,v)$ be an eigenvalue and eigenvector pair for the operator  $E^\top C E$ and let $z \in \RR^{d}$ such that $Ev = Fz$.  We will show that $(\lambda,z)$ is an eigenvalue and eigenvector pair for $F^\top C F$ and thus the proof is complete:
\begin{align*}
\lambda z &= \lambda F^\top F z = \lambda F^\top E v = F^\top E (\lambda v) = F^\top E E^\top C E v = F^\top C F z.
\end{align*}
The last equality follows from the fact that $E E^\top = F F^\top$ is the unique projection matrix onto $\vec{\BB}^n(\Pi)$.
\endproof

\begin{prop}\label{prop:basis_norm}
Let $\Pi \in \BA$. Then $\bigl\|\bigl(\DE\bigr)^{\top} c\bigr\|_2$ is independent of the 
choice of orthonormal basis of perturbation matrices $\vec{E}(\Pi)$ and for any fixed vector $c \in \RR^n$.
\end{prop}
\proof{}
Let $E$ and $F$ be two distinct basis matrices for the vectorized perturbation space $\vec{\BP}^n(\Pi)$ as in the proof of Proposition \ref{prop:basis_eigenvalue}.  By linearity of the directional derivative (see Theorem \ref{thm:derivative}) we can immediately state that $\bigl(\DE\bigr)^{\top} c = E^\top \tilde{c}$ for some vector $\tilde{c} \in \RR^{n^2}$.  Immediately we can see that $\|E^\top \tilde{c}\|_2 = \|F^\top \tilde{c}\|_2$ since $E E^\top = F F^\top$ is the unique projection matrix onto $\vec{\BB}^n(\Pi)$.
\endproof

\end{document}